\documentclass[nofootinbib,aps,superscriptaddress,preprint,floatfix]{revtex4}
\usepackage{epsfig}
\usepackage{setspace}
\usepackage{color}
\usepackage{subfigure}
\usepackage{graphicx}
\usepackage{epstopdf}

\def\beq{\begin{equation}}
\def\eeq{\end{equation}}
\def\bea{\begin{eqnarray}}
\def\eea{\end{eqnarray}}
\def\dslash{\displaystyle{\not}}

\def\chiz{\chi_0}
\def\chih{\chi_{1/2}}
\def\chif{\chi_{1/2}}
\def\ochif{\overline{\chi}_{1/2}}
\def\chifL{{\chi_{1/2}}_{L}}
\def\chifR{{\chi_{1/2}}_{R}}
\def\ochifL{{\overline{\chi}_{1/2}}_{L}}
\def\ochifR{{\overline{\chi}_{1/2}}_{R}}

\def\chiv{\chi_1}

\begin{document}
\vspace{3.0cm}
\preprint{\vbox {\hbox{NSF-KITP-0862}
\hbox{WSU--HEP--1001} 
}}

\vspace*{2cm}

\title{\boldmath Searching for light Dark Matter in heavy meson decays}

\author{Andriy Badin\vspace{5pt}}
\email{a_badin@wayne.edu}
\affiliation{Department of Physics and Astronomy\\[-6pt]
        Wayne State University, Detroit, MI 48201}

\author{Alexey A.\ Petrov\vspace{5pt}}
\email{apetrov@wayne.edu}
\affiliation{Department of Physics and Astronomy\\[-6pt]
        Wayne State University, Detroit, MI 48201}
\affiliation{Michigan Center for Theoretical Physics\\[-6pt]
University of Michigan, Ann Arbor, MI 48109\\[-6pt] $\phantom{}$ }

\begin{abstract}
Beauty and charm $e^+e^-$ factories running at resonance thresholds
have unique capabilities for studies of the production of light Dark
Matter particles in the decays of $B_q (D)$ meson pairs. We
provide a comprehensive study of light Dark Matter production in heavy meson 
decays with missing energy $ \dslash E$ in the final state, such as $B_q (D^0)
\to\dslash E$ and $B_q (D^0) \to \gamma \dslash E$. We argue that such
transitions can be studied at the current flavor factories (and
future super-flavor factories) by tagging the missing-energy decays
with $B(D^0)$ decays ``on the other side."
\end{abstract}

\def\thepage{{}}
\maketitle
\def\thepage{\arabic{page}}

\section{Introduction}

The presence of cold Dark Matter (DM) in our universe provides the
most natural explanation for several observational puzzles, from the
original measurement of the rotational curves~\cite{Zwicky:1933gu}
of galaxies to the observation of background objects in the
Bullet Cluster~\cite{Clowe:2006eq} and spectrum features of the cosmic
microwave background (CMB) fluctuations. In the conventional picture, 
DM accounts for the majority of mass in our Universe. However, the
nature of DM is still very much a mystery, which could intimately
connect astronomical observations with predictions of various
elementary particle theories. Many such theories, with the notable
exception of the Standard Model (SM), predict one or more stable,
electrically-neutral particles in their spectrum~\cite{PrimBH}. These particles
could form all or part of the non-baryonic Dark Matter in the Universe.

Different models provide different assignments for DM particles'
spin and various windows for their masses and couplings to luminous matter. 
In the most popular models DM is a weakly interacting particle particle with mass 
set around the electroweak energy scale. This follows from the
experimental measurements of the relic abundance $\Omega_{DM} h^2
\sim 0.12$ by WMAP collaboration~\cite{Spergel:2003cb} 
\begin{equation}\label{RelAb}
\Omega_{DM} h^2 \sim \langle \sigma_{ann} v_{rel}\rangle^{-1} \propto \frac{M^2}{g^4}
\sim 0.12,
\end{equation}
where $M$ and $g$ are the mass and the interaction strength
associated with DM annihilation respectively. As one can see, 
a weakly-interacting massive particle (WIMP) with electroweak-scale mass naturally
gives the result of Eq.~(\ref{RelAb}). This, coupled with an observation 
that very light DM particles might overclose the Universe  (known as the
Lee-Weinberg limit~\cite{LWL}) , seems to exclude the possibility of
the light-mass solution for DM, setting $M_{DM} > 2-6$~GeV.

A detailed look at this argument reveals that those constraints
could be easily avoided, so even MeV-scale particles can be good DM candidates. 
For instance, DM could be non-fermionic~\cite{Bird:2006jd,Burgess:2000yq}, in which 
case the usual suppression of the DM annihilation cross-section
used in setting the Lee-Weinberg limit does not hold. In addition,
low energy resonances could enhance the cross-section without the
need for a large coupling constant. Other solutions, which also
provide low-mass candidates for DM particles, are also 
possible~\cite{Feng:2008ya,Pospelov:2007mp,Kim:2009ke}.

There are many experiments designed to search for both direct
interactions of DM with the detector and indirect evidence of DM
annihilations in our or other galaxies by looking for the products
such as gamma-rays, positrons and antiprotons. Those can in
principle probe low-mass DM. However, direct searches, performed by
experiments such as DAMA and CDMS~\cite{DAMA}, rely on the measurement of the
kinematic recoil of the nuclei in DM interactions. For cold DM
particles, such measurements lose sensitivity with the decreasing
mass of the WIMP as recoil energy becomes smaller~\cite{Petriello:2008jj}. 
Indirect experiments, such as HESS~\cite{HESS}, are specifically tuned to see large
energy secondaries, only possible for weak-scale WIMPs. The
backgrounds for positron and antiproton searches by HEAT and/or
PAMELA experiments~\cite{PAMELA} could be prohibitively large at small energies. 

It is well-known that the existing $e^+e^-$ flavor factories and
future super-flavor factories could provide the perfect opportunity
to search for rare processes, especially the ones that require high
purity of the final states. In particular, probes of rare B-decays,
such as $B \to K^{(*)} \nu \overline{\nu}$, are only possible at
those machines. These colliders, where $B_q(D)$ and
$\overline{B}_q(\overline{D})$ are produced in charge and
CP-correlated states, have an opportunity to tag the decaying heavy
meson ``on the other side," which provides the charge or
CP-identification of the decaying ``signal" $B$ or $D$ meson. In
fact, many CP-violating parameters at B-factories have been measured
using this method~\cite{Antonelli:2009ws}. It is then possible to perform a similar tag on
the meson decaying to a pair of light DM particles or a pair of DM
particles and a photon. The latter process might become important
for some DM models as it eliminates helicity suppression of the
final state\footnote{ This is similar to the situation in
leptonic decays of $B$-mesons, where the branching ratios ${\cal B}(B \to
\mu\nu\gamma) \approx {\cal B}(B \to \mu\nu)$ and ${\cal B}(B \to e\nu\gamma)
\gg {\cal B}(B \to e\nu)$. }. Moreover, compared to $B \to K + \displaystyle{\not} E$ transitions, 
where $\displaystyle{\not} E$ is missing energy, a massless photon could provide better experimental
opportunities for tagging without reducing the probed parameter space of the DM masses. 
Finally, searches for light DM in heavy meson decays could be more sensitive
than direct detection and other experiments, as DM couplings to heavy quarks
could be enhanced, as for example happens in Higgs portal models~\cite{HiggsPortal}.

In this paper we compute branching ratios for the heavy meson states
decaying into $\chi_s \overline\chi_s$ and $\chi_s \overline\chi_s\gamma$.
Here $\chi_s$ is a DM particle of spin $s$, which appears as missing energy in a
detector. The DM {\it anti-particle} $\overline \chi_s$ may or may not coincide 
with $\chi_s$. We shall first consider model-independent interactions of DM particles 
of spin-0, spin-1/2, and spin-1 with quarks. In each case we write the most general 
effective Hamiltonian coupling DM particles to flavor-changing $b\to q$ (where
$q=s(d)$) or $c \to u$ current and compute $B(D)\to \chi_s
\overline\chi_s (\gamma)$ decay rates. We then consider popular
models, already available in the literature, that can generate
those processes.

\section{Formalism and the Standard Model background}

The computation of decay rates for two-body processes
$B_q(D)\rightarrow \chi_s \overline\chi_s$ is a straightforward task which 
only requires the knowledge of appropriate $B \to \mbox{vacuum}$ matrix 
elements. We use conventional parameterization for those,
\bea\label{CurrMatrEl} \langle 0|\  \overline{b}\gamma^{\mu} q\
|B_q\rangle &=& 0, \qquad \qquad \quad \langle 0|\  \overline{b} q\ |B_q\rangle = 0,
\nonumber \\
\langle 0|\ \overline{b}\gamma^{\mu}\gamma_5 q\ |B_q\rangle &=&i
f_{B_q} P^{\mu}, \qquad \langle 0|\ \overline{b}\gamma_5 q\
|B_q\rangle = -i\frac{f_{B_q} M_{B_q}^2}{m_b + m_q},
\eea
where $P^\mu$ is the 4-momentum of heavy meson $B_q$. Similar formulas 
can be obtained for $D$-meson. In what follows we shall provide relevant derivations for
$B_q$ mesons only, but report results for both $B_q$ and $D^0$-meson 
decays.

Before computing the relevant DM production rates, let us study the
Standard Model background for the decays with missing energy realized in 
transitions to $\nu\overline\nu$ states. The Standard Model effective Hamiltonian for 
$B_q(D) \to \nu\overline{\nu}(\gamma)$ reads
\beq
{\cal H}_{eff}=\frac{4 G_F}{\sqrt{2}}\frac{\alpha}{2\pi\sin^2\theta_W}
\sum_{l=e,\mu,\tau} \sum_{k} \lambda_k X^l(x_k)
\left( J_{Qq}^\mu\right)
\left(\overline{\nu}^l_L \gamma_\mu \nu^l_L\right),
\eeq
where $J_{Qq}^\mu=\overline{q}_L \gamma^\mu b_L$ for beauty, and
$J_{Qq}^\mu=\overline{u}_L \gamma^\mu c_L$ for charm transitions, and we consider
Dirac neutrinos. The functions $\lambda_k X^l(x_k)$ are relevant combinations of the 
Cabbibo-Kobayashi-Maskawa (CKM) factors and Inami-Lim functions. For $b \to q$ transitions 
these functions are overwhelmingly dominated by the top-quark contribution,
\beq
\sum_{k} \lambda_k X^l(x_k) = V_{tq}^* V_{tb} X(x_t), ~\mbox{with}~
X(x_t) = \frac{x_t}{8}\left[\frac{x_t+2}{x_t-1}+\frac{3(x_t-2)}{(x_t-1)^2}\ln{x_t}\right]
\eeq
and $x_t=m_t^2/M_W^2$. Perturbative QCD corrections can be taken into account
by the replacement~\cite{Buchalla:1993bv}
\beq
X_0(x_t) \to \left[X_0(x_t)+\frac{\alpha_s}{4\pi} X_1(x_t)\right]
\left[1-\frac{\alpha_s}{3\pi}\left(\pi^2-\frac{25}{4}\right)\right],
\eeq
where $X_1(x_t)$ can be found in Ref.~~\cite{Buchalla:1993bv}. They change our estimate
by at most 10\%, and therefore be neglected in our analysis. For $c \to u$
transitions we keep the contributions from both internal $b$ and $s$-quarks, so
\beq
\sum_{k} \lambda_k X^l(x_k) = V_{cs}^* V_{us} X^l(x_s)+V_{cb}^* V_{ub} X^l(x_b),
~\mbox{with}~ X^l(x_q) = \overline D(x_q,y_l)/2
\eeq
where $\overline D(x_q,y_l)$ is the Inami-Lim function~\cite{Inami:1980fz} for $y_l=m_l^2/m_W^2$,
\bea
\overline D(x_q,y_l)&=&
\frac{1}{8}\frac{x_q y_l}{x_q-y_l} \left(\frac{y_l-4}{y_l-1}\right)^2 \log y_l +
\frac{1}{8} \left[\frac{x_q}{y_l-x_q}\left(\frac{x_q-4}{x_q-1}\right)^2+1+\frac{3}{(x_q-1)^2}\right]x_q\ln{x_q}\nonumber\\
 &+& \frac{x_q}{4}-\frac{3}{8}\left(1+3\frac{1}{y_l-1}\right)\frac{x_q}{x_q-1}
\eea
Given this, one can easily estimate branching ratios for $B_q(D) \to \nu\overline{\nu}$ decays. One can
immediately notice that the left-handed structure of the Hamiltonian should result in helicity suppression
of those transitions. Assuming for neutrino masses that
$m_\nu \sim \sum_i m_{\nu_i} < 0.62$~eV~\cite{Goobar:2006xz}, where $m_{\nu_i}$ is the mass of one of the
neutrinos, we obtain for the branching ratio
 \begin{equation}
{\cal B}(B_s\rightarrow \nu\overline{\nu})= \frac{G_F^2\alpha^2
f_B^2 M_B^3} {16 \pi^3 \sin^4 \theta_W\Gamma_{B_s}}
|V_{tb}V_{ts}^*|^2X(x_t)^2 x_\nu^2 
\ \simeq  \ 3.07 \times10^{-24}
 \end{equation}
where $x_\nu = m_\nu/M_{B_q}$ and $\Gamma_{B_s}=\Gamma_{B_d}=1/\tau_B$ is 
the total width of the $B_s$ meson. With $\tau_B=1.548$~ps we obtain
${\cal B} (B_d\rightarrow \nu\overline{\nu}) = 1.24\times 10^{-25}$. A
similar calculation yields ${\cal B}(D^0\rightarrow \nu\overline{\nu}) =
1.1 \times10^{-30} $. Clearly such tiny rates imply that decays of
heavy mesons into neutrino-antineutrino final states in the Standard Model can be
safely neglected as sources of background in the searches for DM in
$B_q(D)$-decays. This is one of the main differences between this
study and studies of DM production in $B \to K^{(*)} + \displaystyle{\not} E$ 
transitions~\cite{Bird:2006jd}.

Helicity suppression in the final state can be overcome by adding a third particle, such as a photon, to the
final state. The calculation of $B(D) \to \nu\overline{\nu}\gamma$ has been done before~\cite{Aliev:1996sk}, so here 
we simply present an update. The branching ratio for $B(D) \to \nu\overline{\nu}\gamma$ in principle depends 
on several form-factors,
\bea\label{VAformfactors} \langle \gamma(k)|\overline{b} \gamma_\mu
q| B_q(k+q)\rangle &=&
e~ \epsilon_{\mu\nu\rho\sigma} \epsilon^{*\nu} q^\rho k^\sigma \frac{f_V^B(q^2)}{M_{B_q}}, \nonumber \\
\langle \gamma(k)  |\overline{b} \gamma_\mu \gamma_5 q|
B_q(k+q)\rangle &=& -i e\left[\epsilon^*_\mu \left(kq\right) -
\left( \epsilon^* q\right) k_\mu\right] \frac{f_A^B(q^2)}{M_{B_q}} \eea
\begin{equation}\label{Tformfactors}
\langle \gamma(k)|\overline{b} \sigma_{\mu\nu} q| B_q(k+q)\rangle
=\frac{e}{M_{B_q}^2}\epsilon_{\mu\nu\lambda\sigma}\left[G\epsilon^{*\lambda}k^{\sigma}+H\epsilon^{*\lambda}q^{\sigma}+N(\epsilon^{*}q)q^{\lambda}k^{\sigma}
\right]
\end{equation}
Matrix element $\langle \gamma(k)|\overline{b}
\sigma_{\mu\nu}\gamma_5 q| B_q(k+q)\rangle$ can be obtained using
identity
$\sigma_{\mu\nu}=-\frac{\imath}{2}\epsilon_{\mu\nu\alpha\beta}\sigma_{\alpha\beta}\gamma_5$~\cite{Aliev:2001}
\begin{eqnarray}
G &=& 4g_1, \qquad \qquad \qquad \qquad\quad
N = \frac{-4}{q^2}(f_1+g_1), \nonumber\\
H &=& \frac{-4(qk)}{q^2}\left(f_1+g_1\right), \quad
f_1(g_1) = \frac{f_0(g_0)}{\left(1-q^2/\mu_{f(g)}^2\right)^2}
\end{eqnarray}
where $f_0, g_0,\mu_f, \mu_g$ are known from QCD light-cone sum
rules. Similar formulas hold for $D$-decays. It is important to note that only one 
out of two form-factors is independent. Indeed, as it was shown
in~\cite{Korchemsky:1999qb,Dincer:2001hu},
\beq\label{HQrel} f_V^B(E_\gamma)=f_A^B(E_\gamma) = \frac{f_{B_q}
M_{B_q}}{2 E_\gamma} \left(-Q_q R_q +\frac{Q_b}{m_b}\right)+{\cal
O}\left(\frac{\Lambda_{QCD}^2}{E_\gamma^2}\right)\equiv \frac{f_{B_q}M_{B_q}}{2E_{\gamma}}F_{B_q}, \eeq
where $R_q^{-1}\sim M_{B_q}-m_b$, and $F_{B_q}=
-Q_q R_q +\frac{Q_b}{m_b} \sim \frac{M_{B_q}Q_b-m_b(Q_b+Q_q)}{m_b(M_{B_q}-m_b)}$.
$Q_q=Q_b=+1/3$ are the electrical charges of $q$ and $b$-quarks.
Similar form factor can be obtained for the $D$-meson after a suitable redefinition of quark masses and charges.
One-loop QCD corrections to the Eq.~(\ref{HQrel}) can also be computed~\cite{Lunghi}.

The amplitude for $B_q(D) \to \nu\overline{\nu}\gamma$ transition
could be written as
\bea
A(B_q \to \nu\overline{\nu}\gamma) =
\frac{2 e C_1^{SM}(x_t)}{M_{B_q}} \left[
\epsilon_{\mu\nu\rho\sigma} \epsilon^{*\nu} q^\rho k^\sigma ~f_V^B(q^2) +
i \left[\epsilon^*_\mu \left(kq\right) - \left( \epsilon^* q\right) k_\mu\right]  ~f_A^B(q^2)
\right]  \overline{\nu}_L \gamma^\mu \nu_L,
\eea
where $C_1^{SM}(x_t)=G_F \alpha V_{tb} X_0(x_t)/(2\sqrt{2} \pi
\sin^2\theta_W)$ and $e$ is the electric charge. This results in the
photon spectrum and a branching ratio,
\begin{eqnarray}
\frac{d\Gamma}{dE_{\gamma}} (B_q \to \nu\bar{\nu}\gamma)  &=& \frac{4f_{B_q}^2 G_F^2\alpha^3}{3
M_{B_q}}\left|V_{tb}V_{td}^*X_0(x_t)\right|^2\left(\frac{F_{B_q}}{4
\pi^2\sin^2\theta_W}\right)^2
\nonumber \\
&\times& M_{B_q}^2E_{\gamma}(M_{B_q}+E_{\gamma})\sqrt{\frac{M_{B_q}(1-4x^2)-2E_{\gamma}}{M_{B_q}-2E_{\gamma}}}
\\
{\cal B} (B_q \to \nu\bar{\nu}\gamma)  &=& \frac{2}{\Gamma_{B_q}} f_{B_q}^2 G_F^2\alpha^3 M_{B_q}^5
 \left|V_{tb}V_{td}^*X_0(x_t)\right|^2\left(\frac{F_{B_q}}{12
\pi^2\sin^2\theta_W}\right)^2, 
\end{eqnarray}
where we set $x_\nu=0$. Numerically, 
${\cal B}(B_s\to \nu\bar{\nu}\gamma) = 
{\Gamma (B_s \to \nu\bar{\nu}\gamma)}/{\Gamma_{B_s}} = 3.68\times10^{-8}$. 
Similar results for $B_d$ and $D^0$ mesons are 
${\cal B}(B_d \to \nu\bar{\nu}\gamma) = 1.96\times10^{-9}$ and
${\cal B}(D^0 \to \nu\bar{\nu}\gamma) = 3.96\times 10^{-14}$ respectively.

It is important to notice that the approach to rare radiative
transitions described above works extremely well for SM neutrinos in
the final state since  $E_\gamma \gg \Lambda_{QCD}$ over most of the available 
phase space. It might not be the case for the DM production.
In particular, for $m_{DM} \ge 2$ GeV, the photon energy is quite small
and corrections to Eq.~(\ref{HQrel}) could become significant. Therefore,
our results obtained by using the formalism above should be corrected, for instance, 
using heavy meson chiral techniques.

Currently the only experimental constraints on $B_q (D^0)
\to\dslash E$ and $B_q (D^0) \to \gamma \dslash E$ transitions  are 
available from $B_d$ decays~\cite{Bddata},
\begin{eqnarray}
\label{Bddata}
&&{\cal B}(B_d\to \dslash E) < 2.2\times 10^{-4}, \nonumber\\
&&{\cal B}(B_d\to \dslash E + \gamma) < 4.7\times 10^{-5}.
\end{eqnarray}
One can see that while the branching ratios for the decays into
$\nu\overline{\nu}\gamma$ final states are orders of magnitude
larger than the corresponding decays into $\nu\overline{\nu}$ final
states, they are still way beyond  experimental sensitivities of
currently operating detectors. Thus, we conclude that SM provides no irreducible
background to studies of light DM in such decays.

\section[Heavy meson decaying into pair of bosons and pair of bosons
+ photon]{Scalar Dark Matter production}

\subsection{Generic effective Hamiltonian and $B\rightarrow \chi_0 \overline \chi_0 (\gamma)$ decays}

Let us consider the generic case of a complex neutral scalar field $\chi_0$ describing the DM and limit our
discussion to effective operators of dimensions no more than six.  In this case, a generic effective Hamiltonian
has a very simple form,
\beq \label{ScalLagr} {\cal H}^{(s)}_{eff} =  2 \sum_i
\frac{C_i^{(s)}}{\Lambda^2} O_i, \eeq
where $\Lambda$ is the scale associated with the particle(s) mediating interactions between the
SM and DM fields, and $C_i^{(s)} $ are the Wilson coefficients. The effective operators are
\bea\label{ScalOper}
O_1 &=& m_b (\overline{b}_R q_L)(\chi_0^* \chi_0), \nonumber\\
O_2 &=& m_b (\overline{b}_L q_R)(\chi_0^* \chi_0), \\
O_3 &=&(\overline{b}_L  \gamma^{\mu} q_L)(\chi_0^* \stackrel{\leftrightarrow}{\partial}_{\mu}\chi_0), \nonumber \\
O_4 &=&(\overline{b}_R \gamma^{\mu} q_R)(\chi_0^* \stackrel{\leftrightarrow}{\partial}_{\mu}\chi_0), \nonumber
\eea
where $\stackrel{\leftrightarrow}{\partial} = (\stackrel{\rightarrow}{\partial}-\stackrel{\leftarrow}{\partial})/2$.
For relevant $D$-meson decays one should substitute $m_b \to m_c$ and $b \to q$ currents with $c \to u$
currents. Operators $O_{3,4}$ disappear for DM in the form of real scalar fields.
We note that while the generic form of Eq.~(\ref{ScalOper}) implies that the mediator of interaction between
DM and the SM fields is assumed to be heavy, $M_\Lambda > m_{B_q(D)}$, it is easy to account for the light mediator
by substituting $C_i^{(s)}/\Lambda^2 \to \widetilde{C}_i^{(s)}/(M_{B_q(D)}^2-M_\Lambda^2)$. Clearly, a resonance
enhancement of $B(D) \to \chi_0\chi_0$ rate is possible if for some reason the mediator's mass happens to be
close to $M_{B_q(D)}$. If observed, this resonance enhancement would be seen as anomalously large Wilson
coefficients of the effective Hamiltonian of Eq.~(\ref{ScalOper}).

Let us first compute the $B(D) \to \chi_0 \chi_0$ transition rate. It follows from Eq.~(\ref{ScalOper}) that the 
decay branching ratio is
\beq\label{Bphiphi} 
{\cal B} (B_q \to \chi_0 \chi_0) = 
\frac{\left(C_1^{(s)}-C_2^{(s)}\right)^2}{4\pi M_{B_q} \Gamma_{B_q}}
\left(\frac{f_{B_q}
M_{B_q}^2 m_b}{\Lambda^2 (m_b +m_q)}\right)^2
\sqrt{1-4 x_\chi^2} 
 \eeq
where $x_\chi  = {m_\chi}/{M_{B_q}}$ is a rescaled DM
mass. Clearly, this rate is not helicity-suppressed, so it could be
quite a sensitive tool to determine DM properties at $e^+e^-$ flavor factories. 
The result for a corresponding $D$-decay can
be obtained via trivial substitution of quark masses, widths and
decay constants. Computing the decay rate for various values of Dark
Matter masses and comparing it with the experimental results for
$B_d$ missing energy decays~\cite{Bddata} from Eq.~(\ref{Bddata})
we get the following constraints on coupling constants: 
\begin{eqnarray}
\left(\frac{C_1^{(s)}-C_2^{(s)}}{\Lambda^2}\right)^2&\leq&
2.03\times 10^{-16} ~\mbox{GeV}^{-4} \mbox{ for }
m_{\chi}=0\\
\left(\frac{C_1^{(s)}-C^{(s)}_2}{\Lambda^2}\right)^2&\leq&
2.07\times 10^{-16} ~\mbox{GeV}^{-4} \mbox{ for }
m_{\chi}=0.1\times M_{B_d}\\
\left(\frac{C^{(s)}_1-C^{(s)}_2}{\Lambda^2}\right)^2&\leq&
2.22\times 10^{-16} ~\mbox{GeV}^{-4} \mbox{ for }
m_{\chi}=0.2\times M_{B_d}\\
\left(\frac{C_1^{(s)}-C^{(s)}_2}{\Lambda^2}\right)^2&\leq&
2.54\times 10^{-16} ~\mbox{GeV}^{-4} \mbox{ for }
m_{\chi}=0.3\times M_{B_d} \\
\left(\frac{C^{(s)}_1-C^{(s)}_2}{\Lambda^2}\right)^2&\leq&
3.39\times 10^{-16} ~\mbox{GeV}^{-4} \mbox{ for } m_{\chi}=0.4\times M_{B_d}
\end{eqnarray}
It is worth pointing out that constraints obtained here are much
stricter than those in \cite{Badin}. 

Applying the formalism described above, distribution of the
photon energy and decay width of radiative decay $B_q(D)\rightarrow\chi_0^*\chi_0\gamma$ can be 
computed,
\begin{eqnarray}
\label{dGammaBphiphigamma}
\frac{d\Gamma}{dE_{\gamma}}(B_q \to \chi_0^*\chi_0\gamma)  &=& \frac{f_{B_q}^2
\alpha C^{(s)}_3C^{(s)}_4}{3 \Lambda^4}\left(\frac{F_{B_q}}{4\pi}\right)^2
\frac{2M_{B_q}^2 E_\gamma(M_{B_q}(1-4x_\chi^2)-2E\gamma)^{3/2}}{\sqrt{M_{B_q}-2E\gamma}}\\
\label{GammaBphiphigamma} 
{\cal B} (B_q \to \chi_0^*\chi_0\gamma)&=& \frac{f_{B_q}^2 \alpha
C_3^{(s)}C^{(s)}_4 M_{B_q}^5}{6 \Lambda^4 \Gamma_{B_q}} 
\left(\frac{F_{B_q}}{4\pi}\right)^2 \\
&\times&  \left(\frac{1}{6}\sqrt{1-4x_\chi^2}(1 - 16 x_\chi^2 - 12x_\chi^4)
-12x_\chi^4\log\frac{2 x_\chi}{1+\sqrt{1-4x_\chi^2}} \right),
\nonumber
\end{eqnarray}
We observe that Eqs.~(\ref{dGammaBphiphigamma}) and (\ref{GammaBphiphigamma}) do
not depend on $C^{(s)}_{1,2}$. This can be most easily seen from the
fact that $B_q(D) \to \gamma$ form factors of scalar and pseudoscalar
currents are zero, as follows from Eq.~(\ref{VAformfactors}). Computing decay
rates for various values of Dark Matter mass we are able to restrict
DM properties based on experimental constraints on $B_d$ decays 
with missing energy given in Eq.~(\ref{Bddata}): 
\begin{eqnarray}
&&\frac{C^{(s)}_3}{\Lambda^2}\frac{C^{(s)}_4}{\Lambda^2}
\leq1.55\times10^{-12} ~GeV^{-4} \mbox{\ for\ }m=0\nonumber\\
&&\frac{C^{(s)}_3}{\Lambda^2}\frac{C_4^{(s)}}{\Lambda^2}
\leq1.86\times10^{-12} ~GeV^{-4} \mbox{\ for\ }m=0.1\times M_{B_d}\nonumber\\
&&\frac{C^{(s)}_3}{\Lambda^2}\frac{C_4^{(s)}}{\Lambda^2}
\leq3.20\times10^{-12} ~GeV^{-4} \mbox{\ for\ }m=0.2\times M_{B_d}\\
&&\frac{C_3^{(s)}}{\Lambda^2}\frac{C_4^{(s)}}{\Lambda^2}
\leq9.06\times10^{-12} ~GeV^{-4} \mbox{\ for\ }m=0.3\times M_{B_d}\nonumber\\
&&\frac{C_3^{(s)}}{\Lambda^2}\frac{C_4^{(s)}}{\Lambda^2}
\leq7.44\times10^{-11} ~GeV^{-4} \mbox{\
for\ }m=0.4\times M_{B_d}\nonumber
\end{eqnarray}
Note that Eqs.~(\ref{dGammaBphiphigamma}) and (\ref{GammaBphiphigamma}) depend on $C_3$ and $C_4$, while
Eq.~(\ref{Bphiphi}) only on $C_1$ and $C_2$. Since the models with self-conjugated DM scalar fields only 
contain operators $O_1$ and $O_2$, $B_q(D) \to \chi_0\chi_0(\gamma)$ transitions could be used to test the structure 
of the scalar DM sector. 

\subsection{Production rates in particular models with scalar DM}

In this section we apply the techniques described above for the most general 
effective Hamiltonian for DM particles interacting with the SM fields to
particular model implementations of scalar DM, already available in
the literature. The list of models considered below is by no means exhaustive.

\subsubsection{Minimal and next-to-minimal Scalar Dark Matter models}\label{scalar1}

The simplest possible model for scalar DM involves a real scalar
field $\chi_0\equiv S$ coupled to the SM particles through the
exchange of Higgs boson~\cite{Bird:2006jd,Burges} (see also~\cite{He:2010nt}). 
This is also a very constrained model, where the only two new parameters are the mass
parameter $m_0$ of the scalar DM particle $S$ and the Higgs-scalar
coupling $\lambda$. Nevertheless, it is possible to have light DM in
this model even though it might require some degree of fine-tuning.
The SM Lagrangian is modified by
\begin{eqnarray}\label{MinScalar}
-{\cal L}_S &=& \frac{\lambda_S}{4} S^4 + \frac{m_0^2}{2} S
^2 + \lambda S^2 H^{\dag}H\nonumber\\
& =& \frac{\lambda_S}{4} S^4 + \frac{1}{2}(m_0^2+\lambda
v_{EW}^2) S^2 + \lambda v_{EW} S^2 h +
\frac{\lambda}{2}S^2 h^2
\end{eqnarray}
where $H$ is the Standard Model Higgs doublet, $v_{EW} = 246$~GeV is
the Higgs vacuum expectation value and $h$ is the corresponding
physical Higgs boson. We require $S$ to satisfy $S \to -S$ to make it
a good Dark Matter candidate. The scalar DM particle can be made light by
requiring cancellations between the terms defining its mass,
$m^2=m_0^2+\lambda v_{EW}^2$.

\begin{figure}
\centering
\includegraphics[width=8cm]{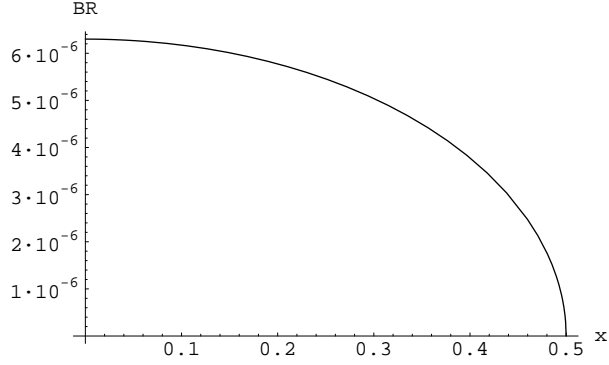}
\caption{${\cal B}(B_d \to SS)$ as a function of $x=m_S/M_{B_d}$. Values of $\lambda$ and $M_h$ were fixed at 1 and $120~GeV$ respectively}
\label{Min:sub} 
\end{figure}

\begin{figure}
\centering
\subfigure[] 
{   \label{MinAllRegLambda:sub:a}
    \includegraphics[width=6cm]{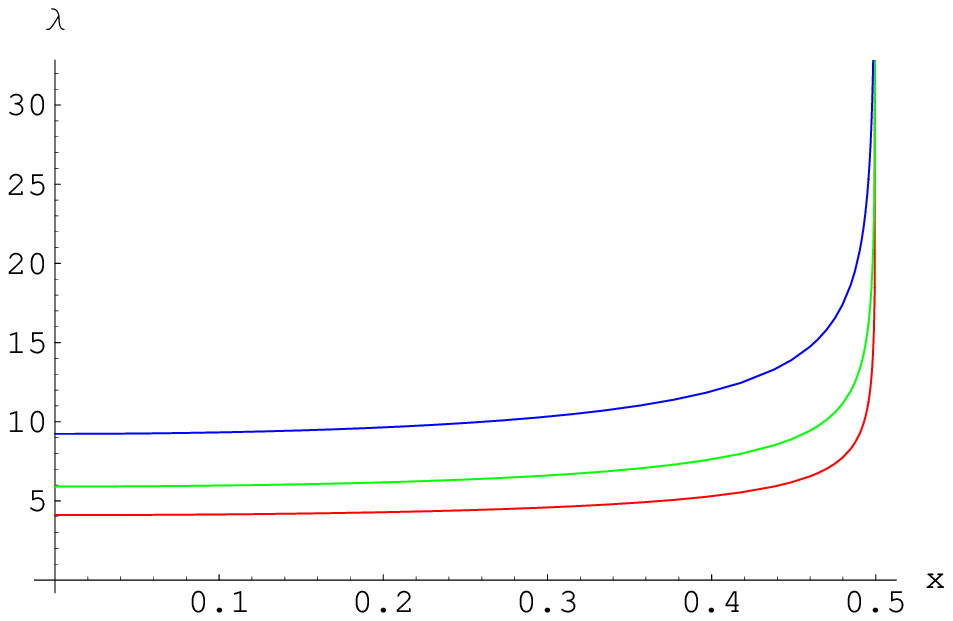}}
\subfigure[] 
{    \label{MinAllRegHiggs:sub:b}
    \includegraphics[width=6cm]{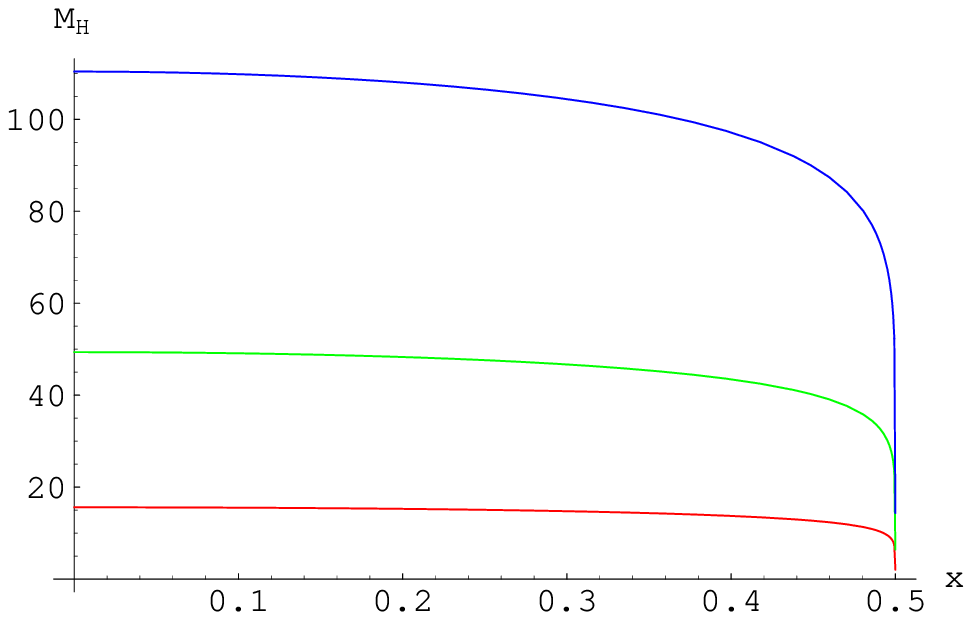}
} \caption{(a) allowed values of the DM-Higgs coupling $\lambda$ as a function of $x=m_S/M_{B_d}$ (below the curves)
for the Higgs masses of 110 GeV (red), 120 GeV (green), and 150 GeV (blue). 
(b) Allowed values of the Higgs mass in GeV (above the curves) for $\lambda=0.1$ (red), 1 (green), and 5 (blue) 
as a function of $x=m_S/M_{B_d}$. 
}
\label{MinAllReg:sub} 
\end{figure}

The transition $B \to SS$ occurs in the minimal model as a one-loop process, 
and since mediating Higgs boson is much heavier than other particles
involved in the process, it can be integrated out. The resulting
effective Hamiltonian reads
\beq\label{MinModHam}
{\cal H}^{(s)}_{eff}=  \frac{3\lambda
g_w^2V_{ts}V_{tb}^*x_t m_b}{64 M_H^2 \pi^2}~ (\overline{b}_L q_R)
S^2, 
\eeq
which implies that $C^{(s)}_{1,3,4}=0$, $C^{(s)}_2 = {3\lambda
g_w^2V_{ts}V_{tb}^*x_t}/{128 \pi^2}$, and $\Lambda = M_H$. Thus,
from Eq.~(\ref{Bphiphi}), the branching ratio for the $B \to SS$ decay in this model is
\beq 
{\cal B} (B_q \to SS)= \left[ \frac{3
g_w^2V_{tq}V_{tb}^*x_t m_b}{128 \pi^2} \right]^2
\frac{\sqrt{1-4x^2_{\rm{S}}}}{16\pi M_B \Gamma_{B_q}}
\left(\frac{\lambda^2}{M_H^4}\right) \left(\frac{f_{B_q}
M_{B_q}^2}{m_b +m_q}\right)^2, 
\eeq
where $x_{\rm{S}}=m_S/m_{B_q}$. Note that this rate depends not
only on the mass of $S$ but also on the parameter
$\kappa=\lambda^2/M_H^4$. This parameter also drives the calculation
of the relic density of $S$~\cite{Burges},
\beq
\sigma_{ann} v_{rel} = \frac{8 v_{EW}^2 \lambda^2}{M_H^2} \times
\lim_{m_{h^*} \to 2 m_S} \frac{\Gamma_{h^*X}}{m_h^*},
\eeq
where $\Gamma_{h^*X}$ is the rate for the decay $h^*\to X$ for a
virtual Higgs with $M_H \sim 2 m_S$. We can, therefore, fix
$\kappa$ from the relic density calculation. This gives for the
branching ratios of $B_q$ and $D$-decays, 
\bea
{\cal{B}}(B_s\to SS) &\approx& \left(4.5 \times 10^{5} \mbox{~GeV}^4\right) \times \frac{\lambda^2}{M_H^4} \sqrt{1-4x_S^2}
\\
{\cal{B}}(B_d\to SS)  &\approx&  \left(1.3 \times 10^{4} \mbox{~GeV}^4\right) \times \frac{\lambda^2}{M_H^4} \sqrt{1-4x_S^2}
\\
{\cal{B}}(D^0\to SS) &\approx& \left( 2.9 \times 10^{-6} \mbox{~GeV}^4\right) \times \frac{\lambda^2}{M_H^4} \sqrt{1-4x_S^2}
\eea
We require the branching ratios to be smaller than the current experimental upper bound~\cite{Bddata} for 
the missing energy decay given in Eq.~(\ref{Bddata}). With this we are able to put the following restriction onto the
parameters of this model:
\begin{equation}
\label{MinimalModelGeneralConstraint} 
\left(\frac{\lambda}{M_H^2}\right)^2\sqrt{1-4x_{S}^2}
\leq 1.68\times10^{-7}.
\end{equation}
We present the resulting branching ratios as a function of $m_{\chiz}$ in Fig.~\ref{Min:sub}. 
Comparing the above branching ratio with the available
experimental data we can put constraints on the
parameters of this model, which we present in Fig.~\ref{MinAllReg:sub}.
%
For the particular values of Dark Matter particles mass we get
\begin{eqnarray}
\left|\frac{\lambda}{M_H^2}\right| &\leq& 8.2\times10^{-4} \mbox{~GeV}^{-2} \mbox{\ for\ }m_S = 0\nonumber\\
\left|\frac{\lambda}{M_H^2}\right| &\leq& 8.3\times10^{-4} \mbox{~GeV}^{-2} \mbox{\ for\ }m_S = 0.1\times M_{B_q} \nonumber\\
\left|\frac{\lambda}{M_H^2}\right| &\leq& 8.6\times10^{-4} \mbox{~GeV}^{-2} \mbox{\ for\ }m_S = 0.2\times M_{B_q}  \\
\left|\frac{\lambda}{M_H^2}\right| &\leq& 9.2\times10^{-4} \mbox{~GeV}^{-2} \mbox{\ for\ }m_S = 0.3\times M_{B_q}  \nonumber\\
\left|\frac{\lambda}{M_H^2}\right| &\leq& 1.1\times10^{-3} \mbox{~GeV}^{-2} \mbox{\ for\ }m_S =0.1\times M_{B_q}  \nonumber
\end{eqnarray}

The minimal scalar model described above can be made
less restricted if we introduce another mediator for DM-SM
interactions, which should somewhat alleviate the fine-tuning
present in the minimal model~\cite{Burges}. This can be done in a
variety of ways. The simplest one is to introduce another Higgs-like field $U$,
\begin{eqnarray}\label{NMinScalar}
\label{scalar} -{\cal L}_{S'} &=& \frac{\lambda_S}{4} S^4 +
\frac{m_0^2}{2} S^2 + (\mu_1 U + \mu_2 U^2) S^2 + V(U) +  \eta' U^2 H^{\dag}H\nonumber\\
& =& \frac{m_s^2}{2} S^2 + \frac{m_u^2}{2}u^2 +  \mu u S^2 +
\eta v_{EW} u h + \ldots,
\end{eqnarray}
where we only display mass and relevant intreaction terms; ellipses stands for 
other terms in the Lagrangian that are irrelevant for this discussion. 

Here $u$ denotes the excitation around vev of $U$, and $\mu$ and
$\eta$ are parameters with values of the order of electroweak scale.
As far as the studies of DM production in heavy flavor decays are
concerned, extended models of this class are equivalent to the
minimal model after suitable redefinition of parameters~\cite{Burges}.
Performing such redefinitions, we obtain
\bea
{\cal{B}}(B_s\to SS) &\approx& \left(2.1\times10^{-4} \right) \times \frac{\eta^2\mu^2}{M_U^4} \sqrt{1-4x_S^2},
\nonumber \\
{\cal{B}}(B_d\to SS) &\approx&  \left(6.3\times10^{-6} \right) \times \frac{\eta^2\mu^2}{M_U^4} \sqrt{1-4x_S^2},
\\
{\cal{B}}(D^0\to SS) &\approx&  \left(1.38\times 10^{-14} \right) \times \frac{\eta^2\mu^2}{M_U^4} \sqrt{1-4x_S^2},
\nonumber 
\eea
where $M_U$ is the mass of the Higgs-like field $U$ of Eq.~(\ref{NMinScalar}).
In the results above, the mass of the Higgs boson was fixed at $M_h = 120$ GeV.
Since the $S$-field is a real scalar field in both the minimal and the extended models,
these models do not give rise to the radiative decay $B_q \rightarrow SS \gamma$.

\subsubsection{Dark Matter with two Higgs doublets (2HDM)}

In this subsection we consider a singlet scalar WIMP $S$ that interacts
with two Higgs doublets, $H_u$ and $H_d$~\cite{Bird:2006jd,He:2008qm},
\begin{eqnarray}
-{\cal L} =\frac{m_0^2}{2} S^2 + \lambda_1 S^2(|H_d^0|^2 +
|H_d^-|^2) + \lambda_2 S^2(|H_u^0|^2 + |H_u^+|^2) +
\lambda_3 S^2(H_d^- H_u^+ - H_d^0 H_u^0).
\end{eqnarray}
We shall assume that $\lambda_1 \gg \lambda_2$, as the opposite
limit gives results that are not different from the minimal
scalar model considered above. The contribution of $\lambda_3$ is 
suppressed because of the cancelation of two diagrams, as explained in~\cite{Bird:2006jd}. 

Calculating the effective Hamiltonian results in the following expressions of 
the Wilson coefficients,
\beq
C^{(s)}_2 = C^{(s)}_1 = \frac{\lambda_1
g_w^2V_{ts}V_{tb}^*x_t(1-a_t+a_t\log{a_t})}{128
\pi^2(1-a_t)^2}\mbox{\ and }\Lambda = M_H,
\eeq
where $a_q = (m_q/M_H)^2$. As in the previous subsection, no decay into the dark
matter with photon possible within the framework of this model. However, decay into a pair of 
dark matter particles is possible 
\bea\label{BtoSSEH}
{\cal B}(B_s \to SS) & \approx &
\left(0.73 \times 10^{2}  \mbox{~GeV}^4\right) \times  \lambda_1^2\sqrt{1-4x_S^2}\left(\frac{a_t\log{a_t}-a_t +
1}{M_H^2(1-a_t)^2}\right)^2,
\nonumber \\
{\cal B}(B_d\rightarrow SS) & \approx &
\left( 2.1 \mbox{~GeV}^4\right) \times
\lambda_1^2\sqrt{1-4x_S^2}\left(\frac{a_t\log{a_t}-a_t +
1}{M_H^2(1-a_t)^2}\right)^2,
\\
{\cal B}(D^0\rightarrow SS) & \approx &
\left( 5.0 \times 10^{2}  \mbox{~GeV}^4\right) \times \lambda_1^2\sqrt{1-4x_S^2}\left(\sum_{q =b,s,d
}V_{uq}V_{cq}^*\frac{a_q\log{a_q}-a_q + 1}{M_H^2(1-a_q)^2}\right)^2.
\nonumber
\eea
Eqs.~(\ref{BtoSSEH}) can be used for constraining parameters of this model
in $B_q\to SS$ transitions.

\section[Heavy meson decaying into pair of fermions and pair of
fermions + photon(meson)]{Fermionic Dark Matter production}

\subsection{Generic effective Hamiltonian and $B_{d(s)}\rightarrow \chif \overline\chif (\gamma)$ decays}

Let us now consider a generic case of fermionic Dark Matter
production. It is possible that the DM particles have
half-integral spin; so many New Physics models, including Minimal
Supersymmetric Standard Model (MSSM), have fermionic DM candidates.
Most of those models, however, naturally assign rather large masses
to their DM candidates. Nevertheless, either after some fine-tuning
of the relevant parameters or after introducing a light DM-SM
mediator, relatively light DM particles are still possible. Let us
consider their production in the decays of heavy mesons. Once again,
limiting ourselves to the operators of dimension of no more than
six, a relevant effective Hamiltonian reads
\begin{equation}\label{FermLagrangian} 
{\cal{H}}_{eff}^{f)} = \frac{4}{\Lambda^2} \sum_i C_i^{(f)} Q_i,
\end{equation}
where $C_i$'s are relevant Wilson coefficient and $\Lambda$
represents the mass scale relevant for DM-quark interactions (e.g. mediator mass). 
In general, there are twelve possible effective operators,
\bea\label{FermOper}
Q_1 &=&(\overline{b}_L\gamma_{\mu}s_L)(\ochifL\gamma^{\mu} \chifL), \qquad
Q_2 = (\overline{b}_L\gamma_{\mu}s_L)\ochifR\gamma^{\mu} \chifR), \nonumber\\
Q_3 &=&(\overline{b}_R\gamma_{\mu}s_R)(\ochifL\gamma^{\mu}\chifL), \qquad
Q_4 = (\overline{b}_R\gamma_{\mu}s_R)(\ochifR\gamma^{\mu}\chifR), \nonumber\\
Q_5 &=&(\overline{b}_L s_R)(\ochifL \chifR), \qquad\qquad
Q_6 = (\overline{b}_L s_R)(\ochifR \chifL),  \\
Q_7 &=&(\overline{b}_R s_L)(\ochifL \chifR), \qquad\qquad
Q_8 = (\overline{b}_R s_L)(\ochifR \chifL), \nonumber\\
Q_9 &=&(\overline{b}_L \sigma_{\mu\nu} s_R)(\ochifL \sigma^{\mu\nu} \chifR), \quad
Q_{10} = (\overline{b}_L \sigma_{\mu\nu} s_R)(\ochifR \sigma^{\mu\nu} \chifL), \nonumber\\
Q_{11} &=&(\overline{b}_R \sigma_{\mu\nu} s_L)(\ochifL \sigma^{\mu\nu} \chifR), \quad
Q_{12} = (\overline{b}_R \sigma_{\mu\nu} s_L)(\ochifR \sigma^{\mu\nu} \chifL), \nonumber
\eea
where the Dark Matter fermion $\chih$ can be either of Dirac or Majorana
type. The latter choice leads to some simplification of the basis.
All needed matrix elements have been given in
Eq.~(\ref{CurrMatrEl}). Note that the matrix elements of the tensor
operators vanish,
\begin{equation}
\langle 0 | \overline{b} \sigma^{\mu \nu} P_{L,R} q |B_q  \rangle =  0.
\end{equation}
For relevant $D$-meson decays one should substitute $m_b \to m_c$
and $b \to q$ currents with $c \to u$ currents. Using the Hamiltonian of 
Eq.~(\ref{FermOper}) we get for the branching ration of $B_q \to \ochif \chif$,
\begin{eqnarray}
\label{Bff} 
{\cal B} (B_q \to \ochif \chif) &=&\frac{f_{B_q}^2 M_{B_q}^3}{16\pi \Gamma_{B_q}\Lambda^2} \sqrt{1-4 x_\chi^2} 
\left[C_{57}C_{68}
\frac{ 4 M_{B_q}^2 x_\chi^2}{(m_b+m_q)^2} - (C_{57}^2+C_{68}^2) \frac{M_{B_q}^2(2x_\chi^2-1)}{(m_b+m_q)^2}
\right. 
\\
&-& 
\left.
2 \widetilde{C}_{1-8} \frac{x_\chi M_{B_q}}{m_b+m_q} + 2(C_{13}+C_{24})^2 x_\chi^2
\right],
\nonumber
\end{eqnarray}
where we employed short-hand notations for the combinations of Wilson coefficients
 $C_{ij}=C_i^{(f)}-C_j^{(f)}$, and $\widetilde{C}_{1-8} = C_{13}C_{57}+C_{24}C_{57}+C_{13}C_{68}+C_{24}C_{68}$.
Due to its larger mass chirality suppression for the GeV-scale Dark Matter is not as severe as for
neutrinos, even for purely left-handed interactions. The obtained result leads to model-independent 
constraints on the Wilson coefficients of Eq.~(\ref{FermLagrangian}), which is based on experimental data for missing energy
decays of $B_d$ meson (see, e.g. Eq.~(\ref{Bddata})). They can be found in Table~\ref{tab:two_body_decay_limits}.
The results presented there can be used to constrain parameters of particular models of fermionic Dark Matter 
considered below.

\begin{table}
\begin{tabular}{|c|c|c|c|c|c|c|c|c|c|c|c|}
\hline\hline

$x_\chi$ & $C_1/\Lambda^2,$ & $C_2/\Lambda^2,$ &
$C_3/\Lambda^2,$ & $C_4/\Lambda^2,$ &
$C_5/\Lambda^2,$ &$C_6/\Lambda^2,$ &
$C_7/\Lambda^2,$ & $C_8/\Lambda^2,$\\

$\phantom{q}$ & $\mbox{GeV}^{-2}$ & $\mbox{GeV}^{-2}$ &
$\mbox{GeV}^{-2}$ & $\mbox{GeV}^{-2}$ &
$\mbox{GeV}^{-2}$ &$\mbox{GeV}^{-2}$ &
$\mbox{GeV}^{-2}$ & $\mbox{GeV}^{-2}$\\

\hline\hline

$ 0 $& -- & -- & -- & -- & $2.3\times10^{-8}$ & $2.3\times10^{-8}$ &$2.3
\times10^{-8}$ &$2.3\times10^{-8}$ \\

$~0.1~$& $~1.9\times10^{-7}$ & $~1.9\times10^{-7}$ &$~1.9
\times10^{-7}$ &$~1.9\times10^{-7}$ & $~2.3\times10^{-8}$& $2.3
\times10^{-8}$& $~2.3\times10^{-8}$& $~2.3\times10^{-8}$ \\

$0.2$& $9.7\times10^{-8}$ & $9.7\times10^{-8}$&$9.7
\times10^{-8}$ &$9.7\times10^{-8}$ & $2.5\times10^{-8}$& $2.5
\times10^{-8}$& $2.5\times10^{-8}$& $2.5\times10^{-8}$ \\

$0.3$& $6.9\times10^{-8}$ & $6.9\times10^{-8}$&$6.9
\times10^{-8}$ & $6.9\times10^{-8}$& $2.8\times10^{-8}$& $2.8
\times10^{-8}$& $2.8\times10^{-8}$& $2.8\times10^{-8}$ \\

$0.4$& $6.0\times10^{-8}$ & $6.0\times10^{-8}$&$6.0
\times10^{-8}$ & $6.0\times10^{-8}$ & $3.6\times10^{-8}$& $3.6
\times10^{-8}$& $3.6\times10^{-8}$& $3.6\times10^{-8}$ \\
\hline \hline
\end{tabular}
\caption{Constraints (upper limits) on the Wilson coefficients of operators of Eq.~(\ref{FermOper}) from 
the $B_q\rightarrow \chif \ochif$ transition. Note that operators $Q_9-Q_{12}$ give no contribution to this decay.}
\label{tab:two_body_decay_limits}
\end{table}

The technique which we use for the computation of $\Gamma({B_q(D)\rightarrow \chif \ochif \gamma})$ is very similar to
the one used for the radiative decay of heavy meson into scalar DM particles discussed above. 
The hadronic part of the matrix element remains the same, we only modify the part that describes Dark Matter.
These lead to
\begin{eqnarray} 
\frac{d\Gamma}{dE_\gamma}&=&\frac{d\Gamma_{1-8}}{dE_{\gamma}}+\frac{d\Gamma_{9-12}}{dE_{\gamma}},
\nonumber\\
\frac{d\Gamma_{1-8}}{dE_{\gamma}}&=&\frac{f_{B_q}^2F_{B_q}^2\alpha M_{B_q}^2E_{\gamma}}{24\pi^2\Lambda^2}\frac{\sqrt{ M_{B_q}(1-4x_\chi^2) -2 E_{\gamma}}}{\sqrt{M_{B_q} -2 E_{\gamma}}}
\nonumber\\
&\times& \left[(C^{2}_1+C^{2}_2+C^{2}_3+C^{2}_4)(M_{B_q} - x_\chi^2 M_{B_q} - E_\gamma) -
(3C_1 C_2 +3C_3 C_4 )x_\chi^2 M_{B_q}\right],
\label{Q1}
\\
\frac{d\Gamma_{9-12}}{dE_{\gamma}}&=&\frac{64\alpha}{3M_{B_q}^2\pi^2\Lambda^2}\left(\frac{E_\gamma ^3}{M_{B_q}-2E_\gamma}\right)\frac{\sqrt{ M_{B_q}(1-4x_\chi^2)-2E_\gamma}}{\sqrt{M_{B_q}-2E_\gamma}}
\nonumber\\
&\times& \left[2\left((C_{10}^{2}+9C_{11} C_{10} -3C_{12} C_{10} +C_{11}^{2}-
3C_{12} C_{11} +3C_{9} (C_{10} +C_{11} +C_{12} ))f_1^2\right.\right.
\nonumber\\
&&-g_1f_1(C_{10}^{2}
+3C_{10} (C_{11} +C_{12} )+C_{11} (C_{11} +3C_{12} )-3C_{9} (C_{10} +C_{11} +C_{12} ))\nonumber\\
&&\left.+2g_1^2(C_{10}^{2} - 6C_{10} C_{11} +
C_{11}^{2})\right)x_\chi^2 M_{B_q}^2
\nonumber\\
&&\left.+(f_1^2-g_1f_1+2g_1^2)(C_{10}^{2} +
C_{11}^{2})(M_{B_q}^2-2M_{B_q}E_\gamma)\right].
\label{Q9}
\end{eqnarray}
While there are many models of light fermionic DM that employ operators $Q_1$ -- $Q_{8}$,
we are not aware of the models with operators $Q_9$ -- $Q_{12}$. Therefore, we chose not to provide a closed 
analytic expression for ${\cal B}_{9-12}(B_q\rightarrow \chif \ochif \gamma)$ here due to overall bulkiness
of the resulting expression.  An interested reader can perform numerical integration of Eq.~(\ref{Q9}) for a 
particular model, if needed. Integrating Eq.~(\ref{Q1}) over the photon energy analytically we obtain 
\begin{eqnarray}
\label{Bffgamma}
{\cal B}_{1-8}(B_q\rightarrow \chif \ochif \gamma) &=&
\frac{F_{B_q}^2f_{B_q}^2 M_{B_q}^2\alpha}{144 \pi^2 \sqrt{1-4x_\chi^2}\Lambda^2}
\nonumber \\
&\times&
\left[\left(C_1^2+C_2^2+C_3^2+C_4^2\right)Y(x_\chi)+ 
\frac{9}{2}\left(C_1C_2 +C_3C_4\right)Z(x_\chi)\right],
\end{eqnarray}
where the factors $Y(x_\chi)$ and $Z(x_\chi)$ are defined as
\bea
\label{StrucFunc}
Y(x_\chi) &=& 1-2x_\chi^2+3x_\chi^2(3-6x_\chi^2+4x_\chi^4)\sqrt{1-4x_\chi^2}\log{\left(\frac{2x_\chi}{1+\sqrt{1-4x_\chi^2}}\right)}-11x_\chi^4+12x_\chi^6,
\nonumber\\
Z(x_\chi) &=& x_\chi^2\left(1+2x_\chi^2 +8x_\chi^2(1-x_\chi^2)\sqrt{1-4x_\chi^2}\log{\left(\frac{2x_\chi}{1+\sqrt{1-4x_\chi^2}}\right)}+8x_\chi^4\right).
\eea
This equation can be used to place constraints on the individual Wilson coefficients of 
Eq.~(\ref{FermOper}). They can be found in Table~\ref{tab:radiative_decay_limits}.
Both Eq.~(\ref{Bff}) and Eq.~(\ref{Bffgamma}) can now be used to constrain the parameters of the 
particular models of fermionic DM.

\begin{table}
\begin{tabular}{|c|c|c|c|c|c|c|c|}
\hline\hline

$x_\chi$ & $~C_1/\Lambda^2,~\mbox{GeV}^{-2}~$ & $~C_2/\Lambda^2,~\mbox{GeV}^{-2}~$ &
$~C_3/\Lambda^2,~\mbox{GeV}^{-2}~$ & $~C_4/\Lambda^2,~\mbox{GeV}^{-2}~$ \\

\hline\hline

$ 0 $& $6.3\times10^{-7}$ &$6.3\times10^{-7}$ &$6.3\times10^{-7}$ &$6.3
\times10^{-7}$ \\

$~0.1~$& $7.0\times10^{-7}$ & $7.0\times10^{-7}$ &$7.0
\times10^{-7}$ &$7.0\times10^{-7}$  \\

$0.2$& $9.2\times10^{-7}$ & $9.2\times10^{-7}$ &$9.2
\times10^{-7}$ &$9.2\times10^{-7}$  \\

$0.3$& $1.5\times10^{-6}$ & $1.5\times10^{-6}$ &$1.5
\times10^{-6}$ &$1.5\times10^{-6}$  \\

$0.4$& $3.4\times10^{-6}$ & $3.4\times10^{-6}$ &$3.4
\times10^{-6}$ &$3.4\times10^{-6}$  \\

\hline\hline
\end{tabular}
\caption{Constraints (upper limits) on the Wilson coefficients of operators of Eq.~(\ref{FermOper}) from 
the $B_q\rightarrow \chif \ochif\gamma$ transition. Note that operators $Q_5-Q_8$ give no contribution to this decay.}
\label{tab:radiative_decay_limits}
\end{table}
%

\subsection{Production rates in particular models with fermionic DM}

\subsubsection{Models with hidden valleys} 

It was pointed out in~\cite{Zurek} that there could be light particles called $v$-quarks interacting with
Standard Model sector via heavy mediator $Z'$. In the simplest
$v$-Model, a $SU(n_v)\times U(1)$ gauge group with couplings $g'$ and
$g_v$ is added to the Standard Model\footnote{The $g'$ coupling constant 
introduced here is not to be confused with the SM hypercharge coupling constant.}. 
The $U(1)$ symmetry is broken by vacuum expectation value of the scalar
field $\langle\phi\rangle$, giving $Z'$ a
mass of $\sim 1-6$~TeV. The latter can mix with Standard Model $Z$
via kinetic mixing $k F^{\mu\nu} F'_{\mu\nu}$. In this model the role of Dark Matter is
played by the $v$-quarks ($\chif\equiv v$).

\begin{figure}
\centering
\subfigure[\ {\cal B} vs m] 
{   \label{HV:BR_vs_m}
    \includegraphics[width=6cm]{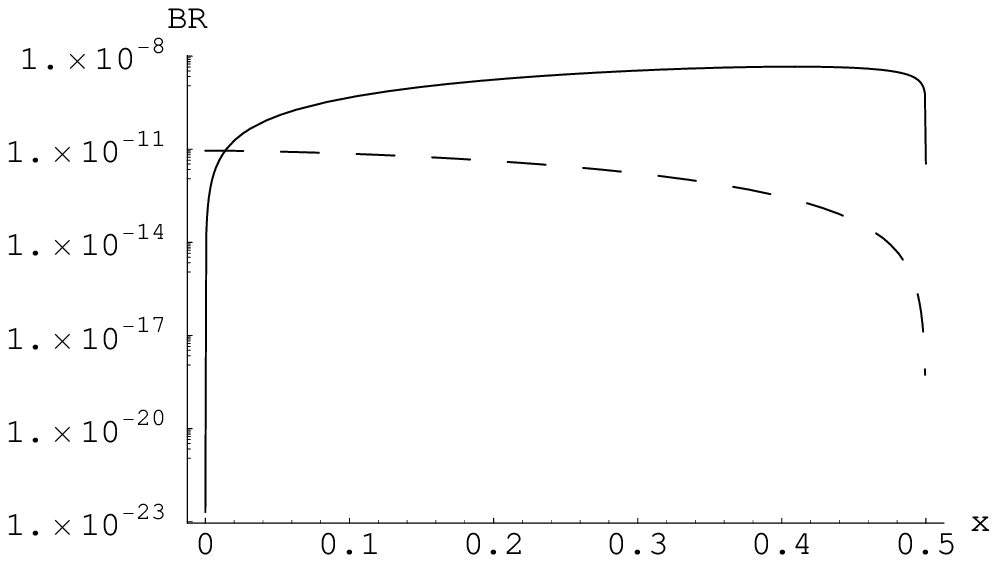}}
\subfigure[] 
{    \label{HV:ar}
    \includegraphics[width=6cm]{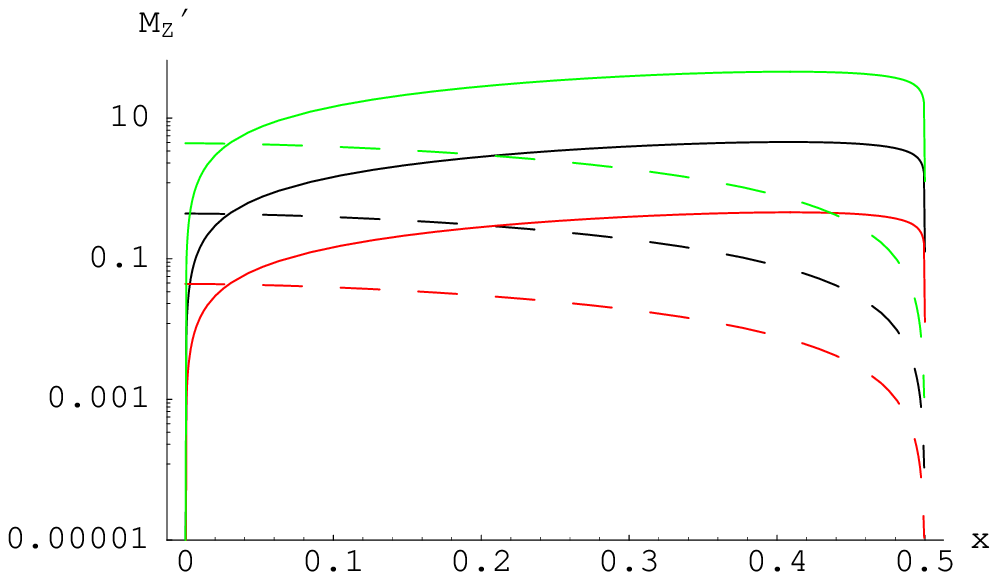}
}
\caption{(a) ${\cal B}(B_d \to vv)$ as a function of $x=m_v/M_{B_d}$ evaluated at $g'=1$, $k=1$ and $M_{Z'}=1~TeV$; 
(b) Allowed values of the $M_{Z^\prime}$ mass in GeV (above the curves) for $g_1 k =1$ (black), 0.1 (red), and 10 (green) as a function of $x=m_v/M_{B_d}$. Solid lines represent the constraints from the 2-body, and
the dashed ones -- from the 3 body (radiative) decay. As one can see, the constraints on the mass of $Z^\prime$ are 
very loose.
}
\label{fig:HV} 
\end{figure}

The model corresponds to the following set of parameters for the
decay of $B_s$ meson (for decays of $B_d$ and $D^0$ parameters will
be similar):
\begin{eqnarray}
C _1 &=& \frac{G_F k g' M_Z M_{Z'}\alpha}{2g_w
\sqrt{2}\sin^2\theta_W}V_{tb}V_{ts}^* X(x), \ \ \mbox{\ and \ }\Lambda = M_{Z'}
\end{eqnarray}
where $k$ is the kinetic mixing parameter, $g^\prime$ is a gauge coupling of the $Z^\prime$ and
$v$-quarks, and $M_{Z^\prime}$ is the mass of the heavy mediator. The rest of
the Wilson coefficients $C_i$ are zero. Thus, from Eq.~(\ref{Bff}),
\begin{equation}
{\cal B}(B_s\to v\overline{v}) \approx (1.76~\mbox{GeV}^2)x^2_v\sqrt{1-4x^2_v}
\left(\frac{ g'k}{M_{Z'}}\right)^2
\end{equation}
where $x_v = m_v/M_{B_q}$. The corresponding results for $B_d$ and $D^0$ decays
are
\begin{equation}
{\cal B}(B_d\to v\overline{v}) \approx (4.68\times10^{-2}~\mbox{GeV}^2) x^2_v\sqrt{1-4x^2_v}\left(\frac{
g'k}{M_{Z'}}\right)^2,
\end{equation}
and 
\begin{equation} 
{\cal B}(D^0\to v\overline{v}) \approx (2.68\times10^{-8} ~\mbox{GeV}^2 ) x^2_v\sqrt{1-4x^2_v} 
\left(\frac{g'k}{M_{Z'}}\right)^2,
\end{equation}
respectively. The corresponding expression for the decay into two $v$-quarks and photon can be obtained 
by defining
\begin{eqnarray}
C_1 = \frac{ G_F k g' \alpha M_Z M_{Z'}}{2g
\sqrt{2}\sin^2\theta_W}V_{tb}V_{ts}^*
X(x)\frac{e}{3}, \ \ \mbox{\ and\ } \Lambda =M_{Z'}.
\end{eqnarray}
We present our results in Fig.~\ref{HV:BR_vs_m} in order to extract the dependence on
DM mass. 
The analytic results for the branching ratios can be well approximated by the following formulas,
\begin{equation}
{\cal B}(B_s\to v\overline{v}\gamma) \approx (2.76\times10^{-4}~\mbox{GeV}^2)\frac{g_1^2 k^2}{M_{Z'}^2}
\times \frac{Y(x_v)}{\sqrt{1-4x_v^2}}
\end{equation}
for the branching ratio of $B_s$ radiative decay and
\begin{eqnarray}
&&{\cal B}(B_d\to v\overline{v}\gamma) \approx (9.07\times10^{-6}~\mbox{GeV}^2)\frac{g_1^2 k^2}{M_{Z'}^2}
\times \frac{Y(x_v)}{\sqrt{1-4x_v^2}}, \\
&&{\cal B}(D^0\to v\overline{v}\gamma) \approx (3.68\times10^{-12}~\mbox{GeV}^2)\frac{g_1^2 k^2}{M_{Z'}^2}
\times \frac{Y(x_v)}{\sqrt{1-4x_v^2}},
\end{eqnarray}
for $B_d$ and $D^0$ decays, respectively. The structure function $Y(x)$ appearing in this 
equation was defined in Eq.~(\ref{StrucFunc}).

\subsubsection{Right-handed massive neutrinos as a Fermionic Dark Matter}

Massive right-handed neutrinos appear naturally in left-right
symmetric models (see for example \cite{Right-Handed-Massive}). The
see-saw mechanism is used to get light left-handed neutrinos and
massive right-handed ones. The coupling of the massive neutrino to
the SM fields in this case is mediated by a right-handed gauge boson
with mass in the TeV range. In this section $\chif \equiv \nu_R$.
\beq {\cal H}_{eff}=\frac{4
G_F^{(R)}}{\sqrt{2}}\frac{\alpha}{2\pi\sin^2\theta_W}
 \sum_{k} \lambda_k X(x_k) \left(
J_{Qq}^\mu\right) \left(\overline{\nu}_R \gamma_\mu \nu_R\right),
\eeq
where $J_{Qq}^\mu=\overline{q}_R \gamma^\mu b_R$ for beauty and
$J_{Qq}^\mu=\overline{u}_R \gamma^\mu c_R$ for charm transitions.
The functions $\lambda_k X(x_k)$ are the combinations of the
Cabbibo-Kobayashi-Maskawa (CKM) factors and Inami-Lim functions.
$G_F^{(R)}$ is defined similarly to the usual Fermi constant,
\begin{equation}
\frac{G_F^{(R)}}{\sqrt{2}} = \frac{g^2}{8 M_{W_R}^2},
\end{equation}
which implies that
\begin{equation}
C_4 = \frac{g^2}{8}\frac{\alpha}{2\pi\sin^2\theta_W} .
\end{equation}
Following the procedure described above, we obtain the following results for decay branching ratios,
\begin{eqnarray}
{\cal B}(B_s\to \nu_R \bar{\nu}_R) &\approx& \frac{3.6\times 10^3 ~\mbox{GeV}^4}{M_{W_R}^4} \ x_v^2\sqrt{1-4x_v^2} \ ,
\\
{\cal B}(B_s\to \nu_R \bar{\nu}_R\gamma) &\approx&\frac{0.57~\mbox{GeV}^4}{M_{W_R}^4}\times Y(x_\nu),
\\
{\cal B}(B_d\to \nu_R \bar{\nu}_R) &\approx& \frac{10^2~\mbox{GeV}^4}{M_{W_R}^4} \ x_v^2\sqrt{1-4x_v^2},
\\
{\cal B}(B_d\to \nu_R \bar{\nu}_R\gamma) &\approx&\frac{1.9\times 10^{-2}~\mbox{GeV}^4}{M_{W_R}^4}\times Y(x_\nu),
\\
{\cal B}(D^0\to \nu_R \bar{\nu}_R) &\approx& \frac{5.6\times 10^{-5}~\mbox{GeV}^4}{M_{W_R}^4} \ x_v^2\sqrt{1-4x_v^2},
\\
{\cal B}(D^0\to \nu_R \bar{\nu}_R\gamma) &\approx&\frac{7.6\times 10^{-9}~\mbox{GeV}^4}{M_{W_R}^4}\times Y(x_\nu),
\end{eqnarray}
where $Y(x)$ is defined in Eq.~(\ref{StrucFunc}).
\begin{figure}
\centering
\subfigure[] 
{   \label{RHN:BR_vs_m}
    \includegraphics[width=6cm]{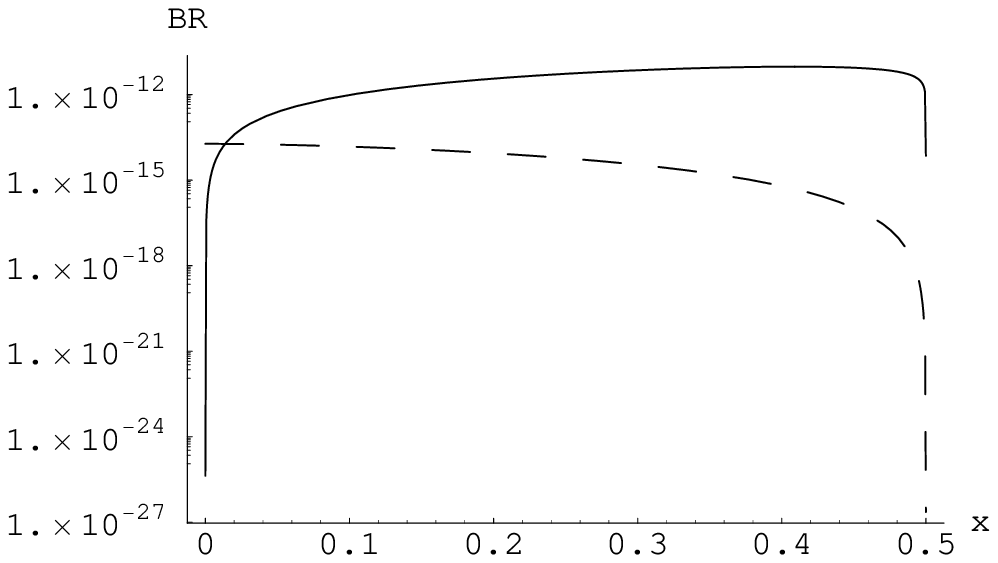}}
\subfigure[] 
{    \label{RHN:ar1}
    \includegraphics[width=6cm]{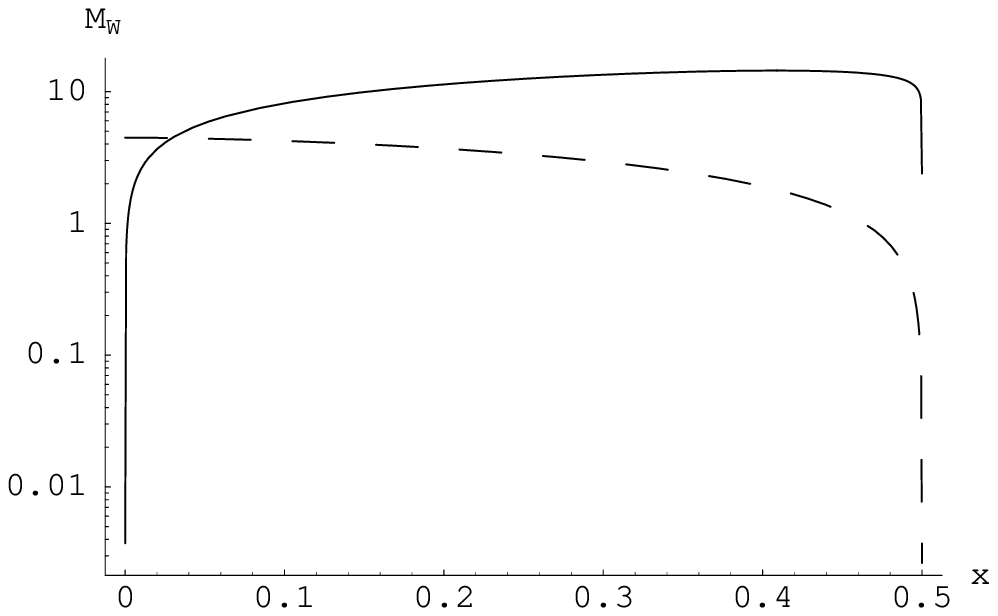}
}
\caption{
(a) ${\cal B}(B_d \to \nu_R\bar \nu_R)$ as a function of $x=m_{\nu_R}/M_{B_d}$ evaluated at $M_{W_R}=1~TeV$, 
(b) Allowed values of the $M_{W_R}$ mass in GeV (above the curves) 
as a function of $x=m_{\nu_R}/M_{B_d}$. Solid lines represent the constraints from the 2-body, and
the dashed ones -- from the 3 body (radiative) decay.  As one can see, the constraints on the mass of $W_R$ are 
very loose.
}
\label{fig:RHN} 
\end{figure}
These results are also presented in Fig.~\ref{fig:RHN}.

\subsection{Majorana fermions}

Majorana particles $\chif \equiv \chi$ often appear in many models of physics beyond the
Standard Model. For generic studies of decays of heavy mesons to Majorana DM particles
we can also use Lagrangian of Eq.~(\ref{FermOper}). The resulting formulas, however,
will be simplified due to the known properties of Majorana fermions~\cite{Haber-Kane},
\begin{eqnarray}\label{MajCon}
\bar{\chi}\gamma_{\mu}\chi &=& 0, \nonumber\\
\bar{\chi}\sigma^{\mu\nu}\chi &=& 0. \nonumber
\end{eqnarray}
Taking into account the conditions of Eq.~(\ref{MajCon}), we can obtain the branching ratio
for $B_q\to \chi\chi$ decay,
\begin{eqnarray}
\label{BffMajorana} 
{\cal B} (B_q \to \chi\chi) &=&\frac{f_{B_q}^2 M_{B_q}^5}{16\pi \Gamma_{B_q} (m_b+m_q)\Lambda^2} \sqrt{1-4 x_\chi^2} 
\left[C_{57}^2 + C_{68}^2 -2x^2 (C_{57}-C_{68})^2\right].
\end{eqnarray}
The photon energy distribution in $B_q\to \chi\chi\gamma$ decay reads
\begin{eqnarray}
\label{GammaBffgammaMajorana}
\frac{d\Gamma}{dE_{\gamma}}&=&\frac{f_{B_q}^2F_{B_q}^2\alpha M_{B_q}^2E_\gamma}{48 \pi^2\Lambda^2}
\frac{\sqrt{M_{B_q}(1-4x_\chi^2)-2E_\gamma}}{\sqrt{M_{B_q}-2E_\gamma}}\times
(C_{12}^2+C^2_{34})(M_{B_q}(1+2x_\chi^2)+E_\gamma),
\end{eqnarray}
which can be integrated over to obtain the branching fraction
\begin{eqnarray}
\label{dGammaBffgammaMajorana}
{\cal B}(B_q \to \chi\chi\gamma)&=&\frac{f_{B_q}^2F_{B_q}^2\alpha M_{B_q}^5}{1152 \pi^2 \Lambda^2}
(C_{12}^2+C^2_{34})\times\\
&&\left(36x_\chi^2 \log{\frac{2x_\chi}{\sqrt{1-4x_\chi^2}+1}}+(4+17x_\chi^2 + 6x_\chi^4)\sqrt{1-4x_\chi^2}\right).
\nonumber
\end{eqnarray}

As an example, we consider a realization of the fermionic dark matter scenario proposed in \cite{Bird:2006jd}.
In this model the Majorana fermion coupled to a higgs-higgsino pair is considered. It must be noted  
that by ``higgsino" we mean a fermionic field with the same quantum numbers as a Higgs field. We, however, do not
place any supersymmetric requirements on the coupling constants. With that,
\begin{eqnarray}
-{\cal L}_f &=& \frac{M}{2} \bar{\psi} \psi + \mu
\bar{\tilde{H_d}} \tilde{H_u} + \lambda_d\bar{\psi}\tilde{H_d}H_d +
\lambda_u\bar{\psi}\tilde{H_u}H_u,
\end{eqnarray}
where $M \ll \mu,\ \lambda_uv_u $. The Dark Matter candidate is the lightest mass eigenstate, which we define as
\begin{eqnarray}
\chi &=& -\psi \cos \theta + \tilde{H_d}\sin\theta, \  \mbox{\ \ \ }
\sin^2 \theta = \frac{\lambda_u^2 v_u^2}{\lambda_u^2 v_u^2 +
\mu^2}\nonumber\\
m_1&=& M \left(1 - \frac{\lambda_u^2 v_u^2}{\lambda_u^2 v_u^2 + \mu^2}
\right).
\nonumber
\end{eqnarray}
We are thus led to the following effective Lagrangian,
\begin{equation}
{\cal{L}}_{eff}=\frac{1}{2}\frac{V_{ts}V_{tb}^*\tan{\beta}}{32\pi^2v_{sm}^3}
\left(\frac{\lambda_d\lambda_u v_u
\mu}{\lambda_u^2v_u^2+\mu^2}\right)\frac{m_b a_t\ln{a_t}}{(1-a_t)} \ (\bar{b}_Ls_R)(\bar{\chi}\chi),
\end{equation}
where $a_t = m_t^2/M_h^2$ and $\tan{\beta} = v_u/v_d$. Matching this Lagrangian to 
Eqs.~(\ref{FermLagrangian}, \ref{FermOper}), we observe that $C _5 = C _6$, and the remaining coefficients $C _i = 0$.
In addition,
\begin{eqnarray}
C _5 = C _6 &=&
\frac{V_{ts}V_{tb}^*\tan{b}}{(16\pi)^2v_{sm}^3}
\left(\frac{\lambda_d\lambda_u v_u
\mu}{\lambda_u^2v_u^2+\mu^2}\right)\frac{m_b m_t^2\ln{a_t}}{(1-a_t)}, \ \mbox{\ and \ }\Lambda = M_h.
\end{eqnarray}
Taking into account  Eq.~(\ref{dGammaBffgammaMajorana}) we conclude that no decay into $\chi\overline{\chi}\gamma$ 
is possible in this particular model. However, a simpler decay into $\chi\chi$
is possible,
\begin{eqnarray}
{\cal B}(B_s\to \chi \chi) &\approx&
1.47\times10^{-10}\sqrt{1 - 4
x_\chi^2}\frac{\log^2(a_t)}{(1-a_t)^2}\left(\frac{\tan(\beta)v_u
\lambda_d\lambda_u\mu}{(v_u^2\lambda_u^2+\mu^2)}\right)^2,
\\
{\cal B} (B_d\to \chi \chi) &\approx&
4.16\times10^{-12}\sqrt{1 -
4x_\chi^2}\frac{\log^2(a_t)}{(1-a_t)^2}\left(\frac{\tan(\beta)v_u
\lambda_d\lambda_u\mu}{(v_u^2\lambda_u^2+\mu^2)}\right)^2,
\\
{\cal B}(D^0\to \chi \chi) &\approx&
1.81\times10^{-11}\sqrt{1 - 4 x_\chi^2}\left(\frac{\tan(\beta)v_u
\lambda_d\lambda_u\mu}{(v_u^2\lambda_u^2+\mu^2)}\sum_{q =
b,\ s,\ d}V_{cq}V_{uq}^*\frac{a_q\log(a_q)}{(1-a_q)}\right)^2,
\end{eqnarray}
where $a_q = (m_q/M_H)^2$ and $x_\chi = m_{\chi}/M_{B_q}$. These results can be used to constrain
the parameters of this model.
\begin{figure}
\centering
\subfigure[] 
{   \label{Majo:BR_vs_m}
    \includegraphics[width=6cm]{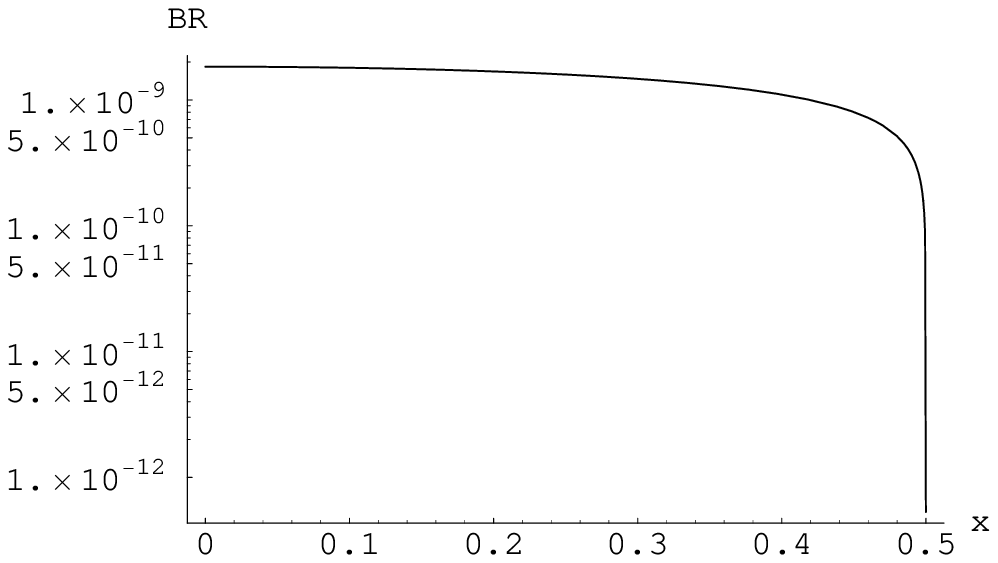}}
\subfigure[] 
{    \label{Majo:ar}
    \includegraphics[width=6cm]{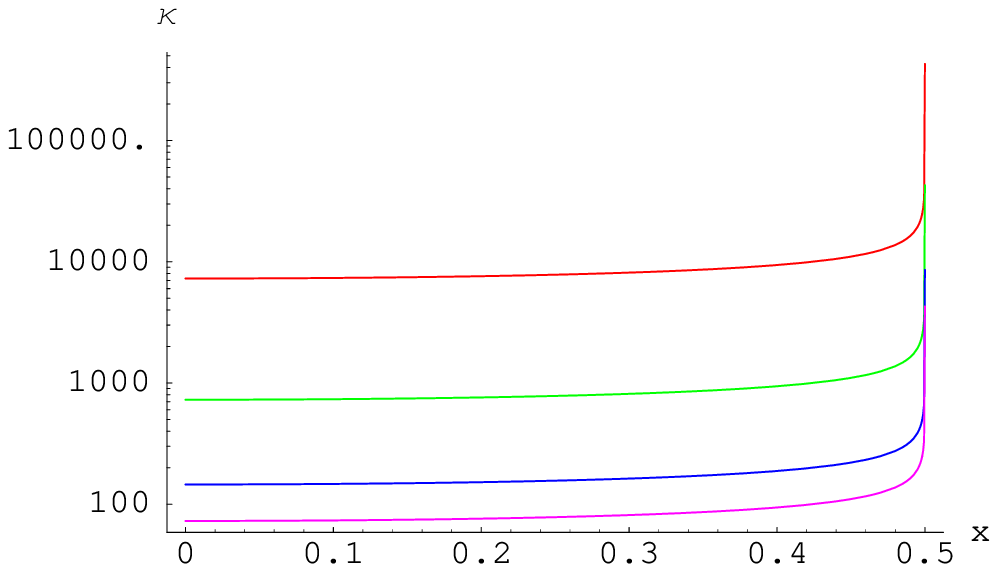}
}
\caption{
(a) ${\cal B}(B_d \to \chi\bar \chi)$ as a function of $x=m_{\chi}/M_{B_d}$. The following numerical values were used: $\kappa = (\lambda_d \lambda_u v_u \mu)/(\lambda_u^2v_u^2+\mu^2) = 1$, $\tan \beta = 10$, $M_h=102~GeV$ 
(b) Allowed values of the $\kappa$ (above the curves) 
for the values of of $\tan\beta=1$  (red), 10 (green), 100 (blue), and 1000 (purple) while mass of Higgs boson was fixed at $M_h=120~GeV$
as a function of $x=m_{\chi}/M_{B_d}$.  
}
\label{fig:Majo} 
\end{figure}
%

\section{Vector Dark Matter production. Generic effective Hamiltonian and $B_q (D^0) \to \chiv\chiv$ decays}

Vector DM is a quite popular concept in non-supersymmetric solutions of the hierarchy problem.
In particular, it can be encountered in models with Universal Extra Dimensions (UED), little Higgs models with
T-parity, and some variations of Randall-Sundrum models. All of the proposed models that the
authors are aware of involve weak-scale DM particles. This however, does not preclude the
existence of the low mass vector DM.

Let us consider a generic case of a vector field $\chiv^\mu$ describing
Dark Matter. This DM particle could be either a gauge boson,
corresponding to some abelian or non-abelian gauge symmetry broken
at some higher scale, or some composite state. The only assumption
that we shall make is that $\chiv$ is odd under some $Z_2$-type discrete
symmetry, $\chiv^\mu \to - \chiv^\mu$. This condition results in the
pair-production of DM particles.

We shall limit our discussion to the effective operators of the
dimension no more than six.  Since no gauge symmetry related to
$\chiv^\mu$ is present at the scale $m_Q$, the most general effective
Hamiltonian should be built out of the vector field $\chiv^\mu$ and its
field strength tensor $\chiv^{\mu\nu}$. In this case, an effective
Hamiltonian has a very simple form,
\beq\label{VecHam}
{\cal H}^{(v)}_{eff} =  \sum_i \frac{C_i^{(v)}}{\Lambda^2} O_i,
\eeq
where $\Lambda$ is the scale associated with the mass of the particle mediating interactions between the
SM and DM fields, and $C_i^{(V)} $ are the Wilson coefficients. The effective operators are
\bea\label{VecOper}
O_1 &=& m_b (\overline{b}_L q_R) {\chiv}_\mu \chiv^\mu,
\qquad
O_4 =(\overline{b}_R \gamma_{\mu} q_R) \chiv^{\mu\nu} {\chiv}_\nu,
\nonumber \\
O_2 &=& m_b (\overline{b}_R q_L) {\chiv}_\mu \chiv^\mu,
\qquad
O_5 = (\overline{b}_L  \gamma_{\mu} q_L) \widetilde \chiv^{\mu\nu} {\chiv}_\nu,
\\
O_3 &=&(\overline{b}_L  \gamma_{\mu} q_L) \chiv^{\mu\nu} {\chiv}_\nu,
\qquad
O_6 =(\overline{b}_R \gamma_{\mu} q_R) \widetilde \chiv^{\mu\nu} {\chiv}_\nu, \nonumber
\eea
where $\widetilde \chiv^{\mu\nu} = (1/2) \epsilon^{\mu\nu\alpha\beta}
{\chiv}_{\alpha\beta}$ and $q=s,d$. As before, the Hamiltonian relevant
for charmed meson decays can be obtained by the proper substitution
of $b\to q$ current with $c\to u$ current.

The $B_q(D) \to \chiv\chiv$ transition rate can be computed using Eq.~(\ref{VecOper}). Using the form-factors defined 
in Eq.~(\ref{CurrMatrEl}), we obtain
\bea\label{VecDecay}
{\mbox B}(B_q \to \chiv\chiv) &=&
\frac{f_B^2 M m_b^2 \sqrt{M^4 \left(1-4 x_\chi^2\right)}}{256 (m_b
+m_q)^2 \pi x_\chi^4 \Gamma_{B_q}\Lambda^4} \left[ C^{2}_{12} \left(1-4 x_
\chi^2+12 x_\chi^4 \right)\right.
\nonumber \\
&+&\left. (m_b+m_q)^2
\left(8 C_{56}^2 \left(1-4 x_\chi^2\right)+3 C_{34}^2\right) x_\chi^4 \right.
\\
&+& 
\left. 2 C_{12}C_{34}
(m_b+m_q) \left(1+2 x_\chi^2\right) x_\chi^2\right],
\nonumber
\eea
where $C_{ik}=C_i^{(v)}-C_k^{(v)}$ and $x_{\rm{DM}} = m_{\chi}/m_{B_q}$.
It is necessary to point out that Eq.~(\ref{VecDecay}) is divergent
at $m_\chi=0$, which is related to the fact that operators in Eq.~(\ref{VecOper})
contributing to the effective Lagrangian are not gauge invariant. Thus, for the case of 
massless DM the upper limit on the Wilson coefficients $C^{(v)}_1$ and $C^{(v)}_2$ is 
zero (see Table \ref{tab:radiative_decay_limits_V}).

Using Eq.~(\ref{VecDecay}), we can place general constraints on the Wilson coefficients of the effective 
Hamiltonian describing interactions of vector DM with quarks (see Eq.~(\ref{VecHam})). They are
presented in Table~\ref{tab:radiative_decay_limits_V}.

We are not aware of particular models of light DM with spin-1 particles and masses 
$m_{\chi} < 3$~GeV.

\begin{table}
\begin{tabular}{|c|c|c|c|c|c|c|c|c|c|}
\hline\hline

$x_\chi$ & $C_1/\Lambda^2,~\mbox{GeV}^{-2}$ & $C_2/\Lambda^2,~\mbox{GeV}^{-2}$ &
$C_3/\Lambda^2,~\mbox{GeV}^{-2}$ & $C_4/\Lambda^2,~\mbox{GeV}^{-2}$ &
$C_5/\Lambda^2,~\mbox{GeV}^{-2}$ &$C_6/\Lambda^2,~\mbox{GeV}^{-2}$ \\

\hline\hline

$ 0 $& $0$ &$0$ &$1.4\times10^{-8}$ &$1.4\times10^{-8}$ &$8.9
\times10^{-9}$ &$8.9\times10^{-9}$ \\

$~0.1~$& $1.2\times10^{-9}$ & $1.2\times10^{-9}$ &$1.5
\times10^{-8}$ &$1.5\times10^{-8}$ &$9.1\times10^{-9}$ & $9.1
\times10^{-9}$ \\

$0.2$& $5.1\times10^{-9}$ & $5.1\times10^{-9}$ &$1.5
\times10^{-8}$ &$1.5\times10^{-8}$ &$1.0\times10^{-8}$ &$1.0
\times10^{-8}$ \\

$0.3$& $1.3\times10^{-8}$ & $1.3\times10^{-8}$ &$1.6
\times10^{-8}$ &$1.6\times10^{-8}$ & $1.2\times10^{-8}$&$1.2
\times10^{-8}$ \\

$0.4$& $2.9\times10^{-8}$ & $2.9\times10^{-8}$ &$1.9
\times10^{-8}$ &$1.9\times10^{-8}$ & $1.9\times10^{-8}$&$1.9
\times10^{-8}$ \\

\hline\hline

\end{tabular}
\caption{Constraints (upper limits) on the Wilson coefficients of operators of Eq.~(\ref{VecOper}) from 
the $B_q\rightarrow \chiv \chiv$ transition.}
\label{tab:radiative_decay_limits_V}
\end{table}
%

\section{Conclusions}\label{Conclusions}

We have argued that missing energy decays of the heavy mesons -
$B_d$, $B_s$ and $D^0$ - provide an important way to probe different
properties of Dark Matter. Consideration of different decay modes -
two body decays, radiative and light meson + DM decays -  restricts
different regions of the Dark Matter parameter space. Combined
constraints obtained from different decay modes of various heavy
mesons provide indispensable probe of physics beyond the Standard
Model in general and the nature of the Dark Matter in particular.
For instance, observation of $B_q (D^0) \to \gamma \dslash E$, but 
non-observation of $B_q (D^0)\to\dslash E$ transitions directly point
to non-self-conjugated nature of scalar DM. 

We reported general constraints on the Wilson coefficients of the 
effective operators describing interactions of DM with quarks (see Tables I-III).
Restrictions obtained in our paper are much stricter than
constraints from single decay modes. Our results combined with
constraints from astrophysical observables (for example
\cite{Badin}), direct detection of Dark Matter and
invisible decays of heavy hadrons \cite{Gagik} could provide a full set of
tools needed to test (or rule out) the models of light Dark Matter.

\section{Acknowledgements}

We would like to thank B.~McErlath and S.~Blusk for useful conversations,
and A.~Blechman for carefully reading the manuscript and useful suggestions.
We also thank J.~Tandean for catching a misprint in the first version of this paper.
This work was supported in part by the U.S.\ National Science
Foundation under CAREER Award PHY--0547794, and by the U.S.\
Department of Energy under Contract DE-FG02-96ER41005. A.A.P. would
like to thank the Kavli Institute for Theoretical Physics (KITP) at
UCSB for hospitality and acknowledge support for his stay at KITP
from the National Science Foundation under Grant PHY05-51164. A.A.P.
also thanks CERN Theory Division and Galileo Galilei Institute for Theoretical Physics in
Florence (Italy) where part of this work was done.




\newpage
\begin{thebibliography}{99}

\bibitem{Zwicky:1933gu}
  F.~Zwicky,
  Helv.\ Phys.\ Acta {\bf 6}, 110 (1933).

\bibitem{Clowe:2006eq}
  D.~Clowe, M.~Bradac, A.~H.~Gonzalez, M.~Markevitch, S.~W.~Randall, C.~Jones and D.~Zaritsky,
  Astrophys.\ J.\  {\bf 648}, L109 (2006)
  [arXiv:astro-ph/0608407].

\bibitem{PrimBH}
While no elementary particle with appropriate for DM properties exist in the SM, it does not mean that 
New Physics is required to explain DM. For example, DM could exist in the form of small primordial black holes. 
For more details see  C.~Bambi, A.~D.~Dolgov and A.~A.~Petrov,
  Phys.\ Lett.\  B {\bf 670}, 174 (2008)
  [Erratum-ibid.\  {\bf 681}, 504 (2009)]
  [arXiv:0801.2786 [astro-ph]];
  P.~H.~Frampton, M.~Kawasaki, F.~Takahashi and T.~T.~Yanagida,
  JCAP {\bf 1004}, 023 (2010)
  [arXiv:1001.2308 [hep-ph]];
    A.~D.~Dolgov, P.~D.~Naselsky and I.~D.~Novikov,
  arXiv:astro-ph/0009407.

\bibitem{Spergel:2003cb}
  D.~N.~Spergel {\it et al.}  [WMAP Collaboration],
  Astrophys.\ J.\ Suppl.\  {\bf 148}, 175 (2003)
  [arXiv:astro-ph/0302209].

\bibitem{LWL}
  B.~W.~Lee and S.~Weinberg,
  Phys.\ Rev.\ Lett.\  {\bf 39}, 165 (1977);
  M.~I.~Vysotsky, A.~D.~Dolgov and Y.~B.~Zeldovich,
  JETP Lett.\  {\bf 26}, 188 (1977)
  [Pisma Zh.\ Eksp.\ Teor.\ Fiz.\  {\bf 26}, 200 (1977)];
  for the particular case of light neutralinos, see 
    A.~Bottino, N.~Fornengo and S.~Scopel,
  Phys.\ Rev.\  D {\bf 67}, 063519 (2003)
  [arXiv:hep-ph/0212379].


\bibitem{Bird:2006jd}
  C.~Bird, R.~Kowalewski and M.~Pospelov,
  Mod.\ Phys.\ Lett.\  A {\bf 21}, 457 (2006)

\bibitem{Burgess:2000yq}
  C.~P.~Burgess, M.~Pospelov and T.~ter Veldhuis,
  Nucl.\ Phys.\  B {\bf 619}, 709 (2001)
  [arXiv:hep-ph/0011335].


\bibitem{Feng:2008ya}
  J.~L.~Feng and J.~Kumar,
  Phys.\ Rev.\ Lett.\  {\bf 101}, 231301 (2008)
  [arXiv:0803.4196 [hep-ph]].

\bibitem{Pospelov:2007mp}
  M.~Pospelov, A.~Ritz and M.~B.~Voloshin,
  Phys.\ Lett.\  B {\bf 662}, 53 (2008)
  [arXiv:0711.4866 [hep-ph]].

\bibitem{Kim:2009ke}
  Y.~G.~Kim and S.~Shin,
  JHEP {\bf 0905}, 036 (2009)
  [arXiv:0901.2609 [hep-ph]].

\bibitem{DAMA} DAMA collaboration website: http://people.roma2.infn.it/~dama/web/home.html;
CDMS collaboration website: http://cdms.berkeley.edu/;
for a recent example of CDMS/DAMA sensitivity to light DM candidates in SUSY models see
  A.~Bottino, F.~Donato, N.~Fornengo and S.~Scopel,
  arXiv:0912.4025 [hep-ph].


\bibitem{Petriello:2008jj}
  F.~Petriello and K.~M.~Zurek,
  JHEP {\bf 0809}, 047 (2008)
  [arXiv:0806.3989 [hep-ph]].
  
\bibitem{HESS} HESS collaboration website: http://www.mpi-hd.mpg.de/hfm/HESS/

\bibitem{PAMELA} PAMELA collaboration website:  http://pamela.roma2.infn.it/index.php;
see also S. W. Barwick et al. [HEAT Collaboration], Astrophys. J. 482, L191 (1997) [arXiv:astro-ph/9703192];
 S. Coutu et al., Astropart. Phys. 11, 429 (1999), [arXiv:astro-ph/9902162].

\bibitem{Antonelli:2009ws}
  M.~Antonelli {\it et al.},
  arXiv:0907.5386 [hep-ph];
  for charm physics observables see 
    D.~Atwood and A.~A.~Petrov,
  Phys.\ Rev.\  D {\bf 71}, 054032 (2005)
  [arXiv:hep-ph/0207165];
  M.~Artuso, B.~Meadows and A.~A.~Petrov,
  Ann.\ Rev.\ Nucl.\ Part.\ Sci.\  {\bf 58}, 249 (2008)
  [arXiv:0802.2934 [hep-ph]];
  A.~A.~Petrov,
{\it In the Proceedings of Flavor Physics and CP Violation (FPCP 2003), Paris, France, 3-6 Jun 2003, pp MEC05}
  [arXiv:hep-ph/0311371].

\bibitem{HiggsPortal}
  For a recent review and references see, for example, 
  M.~H.~G.~Tytgat,
  arXiv:0906.1100 [hep-ph].

\bibitem{Buchalla:1993bv}
  G.~Buchalla and A.~J.~Buras,
  Nucl.\ Phys.\  B {\bf 400}, 225 (1993).

 \bibitem{Inami:1980fz}
  T.~Inami and C.~S.~Lim,
  Prog.\ Theor.\ Phys.\  {\bf 65}, 297 (1981)
  [Erratum-ibid.\  {\bf 65}, 1772 (1981)].

\bibitem{Goobar:2006xz}
  A.~Goobar, S.~Hannestad, E.~Mortsell and H.~Tu,
  JCAP {\bf 0606}, 019 (2006)
  [arXiv:astro-ph/0602155].

\bibitem{Aliev:1996sk}
  T.~M.~Aliev, A.~Ozpineci and M.~Savci,
  Phys.\ Lett.\  B {\bf 393}, 143 (1997)
  [arXiv:hep-ph/9610255];
  C.~D.~Lu and D.~X.~Zhang,
  Phys.\ Lett.\  B {\bf 381}, 348 (1996)
  [arXiv:hep-ph/9604378].

\bibitem{Aliev:2001}
  T.~M.~Aliev, A.~\"{O}zpineci and M.~Savci
   Phys.Lett.B520:69-77,2001
  [arXiv:hep-ph/0105279 ];
  T.~M.~Aliev and C.~S.~Kim,
  Phys.\ Rev.\  D {\bf 58}, 013003 (1998)
  [arXiv:hep-ph/9710428].


\bibitem{Korchemsky:1999qb}
  G.~P.~Korchemsky, D.~Pirjol and T.~M.~Yan,
  Phys.\ Rev.\  D {\bf 61}, 114510 (2000)
  [arXiv:hep-ph/9911427].

\bibitem{Dincer:2001hu}
  Y.~Dincer and L.~M.~Sehgal,
  Phys.\ Lett.\  B {\bf 521}, 7 (2001)
  [arXiv:hep-ph/0108144].

\bibitem{Lunghi} 
E. Lunghi, D. Pirjol, and D. Wyler, Nucl.
Phys. B 649 (2003) 349-364, C.W. Bauer, S. Fleming, M. Luke, Phys.
Rev. D 63 (2001) 014006, C.W. Bauer, S. Fleming, D. Pirjol, I.W.
Stewart, Phys. Rev. D 63 (2001) 114020, C.W. Bauer, I.W. Stewart, Phys.
Lett. B 516 (2001) 134, C.W. Bauer, D. Pirjol, I.W. Stewart, Phys.
Rev. D 65 (2002) 054022
  
\bibitem{Bddata} 
C. Amsler et al. (Particle Data Group), Physics Letters B667, 1 (2008);
BABAR Collaboration, B.~Aubert et al., Phys.Rev.Lett.93:091802,2004 [arXiv: hep-ex/0405071].

\bibitem{Badin}   
  A.~Badin, G.~K.~Yeghiyan and A.~A.~Petrov,
  arXiv:0909.5219 [hep-ph].

\bibitem{Burges} 
V. Silveira and A. Zee, Phys. Let. B161,
136(1985); J. McDonald Phys. Rev. D50, 3637 (1994)


\bibitem{He:2010nt}
  X.~G.~He, S.~Y.~Ho, J.~Tandean and H.~C.~Tsai,
  arXiv:1004.3464 [hep-ph].

\bibitem{He:2008qm}
  X.~G.~He, T.~Li, X.~Q.~Li, J.~Tandean and H.~C.~Tsai,
  Phys.\ Rev.\  D {\bf 79}, 023521 (2009)
  [arXiv:0811.0658 [hep-ph]].

\bibitem{Zurek} 
  M.~J.~Strassler and K.~M.~Zurek,
  Phys.\ Lett.\  B {\bf 651}, 374 (2007)
  [arXiv:hep-ph/0604261].

\bibitem{Right-Handed-Massive} J.-M. Fr\'{e}re et al., Phys.Rev.D75:085017, 2007 [arxiv:hep-ph/0610240v2]

\bibitem{Haber-Kane} 
H. E. Haber, G. L. Kane
Phys. Rept. 117 (1985) 75-263.

\bibitem{Gagik} 
  G.~K.~Yeghiyan,
  Phys.\ Rev.\  D {\bf 80}, 115019 (2009)
  [arXiv:0909.4919 [hep-ph]].


\end{thebibliography}
\end{document}